\documentclass[aps,prl,10pt,a4paper,twocolumn,superscriptaddress,citeautoscript,showpacs,longbibliography,footinbib]{revtex4}

\usepackage{graphicx}
\usepackage{appendix}
\usepackage{bibunits}
\usepackage{multirow}
\usepackage{color}
\usepackage{bm}
\usepackage{times}
\usepackage{amsmath,bm,amsfonts}
\usepackage{dcolumn}
\usepackage{graphicx}
\usepackage{latexsym}
\usepackage{BOONDOX-cal}
\usepackage{ulem} 
\usepackage{braket}
\usepackage{mathtools}
\usepackage{cancel}

\begin{document}

\title{Phonon angular momentum Hall effect}

\author{Sungjoon \surname{Park}}

\affiliation{Center for Correlated Electron Systems, Institute for Basic Science (IBS), Seoul 08826, Korea}

 \affiliation{Department of Physics and Astronomy, Seoul National University, Seoul 08826, Korea}
 
\affiliation{Center for Theoretical Physics (CTP), Seoul National University, Seoul 08826, Korea}

\author{Bohm-Jung \surname{Yang}}
\email{bjyang@snu.ac.kr}

\affiliation{Center for Correlated Electron Systems, Institute for Basic Science (IBS), Seoul 08826, Korea}

 \affiliation{Department of Physics and Astronomy, Seoul National University, Seoul 08826, Korea}

\affiliation{Center for Theoretical Physics (CTP), Seoul National University, Seoul 08826, Korea}

\date{\today}

\begin{abstract}

Spin Hall effect is the transverse flow of the electron spin in conductors under external electric field. 
Similarly, thermal gradient in magnetic insulators can drive a transverse flow of the spin angular momentum of magnons, which provides a thermal alternative for spin manipulation.
Recently, the phonon angular momentum (PAM), which is the angular momentum of atoms as a result of their orbital motion around their equilibrium positions, has garnered attention as a quantity analogous to the magnon spin.
However, can we manipulate PAM like magnon spin?
Here, we show that temperature gradient generally induces a transverse flow of PAM, which we term the phonon angular momentum Hall effect (PAMHE). 
The PAMHE relies only on the presence of transverse and longitudinal acoustic phonons, and it is therefore ubiquitous in condensed matter systems.
As a consequence of the PAMHE, PAM accumulates at the edges of a crystal.
When the atoms in the crystal carry nonzero Born effective charge, the edge PAM induces edge magnetization, which may be observed through optical measurement.
We believe that PAMHE provides a new principle for the manipulation of angular momenta in insulators and opens up an avenue for developing functional materials based on phonon engineering. 
\end{abstract}

\maketitle
\begin{bibunit}
Transverse responses of materials to external forces, generally known as Hall effects, have played quintessential roles in advancing fundamental physics as well as technology~\cite{nagaosa2010anomalous,sinova2015spin}.
For instance, when external electric field is applied to a conductor, electric current can flow in the transverse direction, which is known as the anomalous Hall effect  \cite{karplus1954hall,smit1958spontaneous,berger1970side,nagaosa2010anomalous}.
Similarly, spin and orbital angular momenta can flow in the direction transverse to the electric field, leading to the spin \cite{d1971possibility,hirsch1999spin,murakami2003dissipationless,sinova2004universal,kato2004observation,wunderlich2005experimental,sinova2015spin} and orbital Hall effects \cite{bernevig2005orbitronics}, respectively.
The intrinsic mechanisms for the various Hall effects \cite{karplus1954hall,murakami2003dissipationless,sinova2004universal,bernevig2005orbitronics} are of particular interest as they do not rely on scattering mechanisms and are closely related to topological phases such as Chern insulators \cite{haldane1988model} and quantum spin Hall insulators \cite{kane2005quantum,kane2005z}.
Moreover, the manipulation of the charge, spin, and orbital degrees of freedom by Hall effects facilitates efficient device engineering based on functional materials \cite{popovic1989hall,jungwirth2012spin}.

Hall effects can occur not only in conductors but also in insulators, even though they do not respond well to electric field.
Here, the key idea is that temperature gradient can apply a statistical force to the quasiparticles in insulators, such as magnon and phonon, in a manner analogous to the electric field in a conductor.
For example, when a temperature gradient is applied, heat current can flow in the transverse direction to the temperature gradient through the low-energy charge-neutral excitations such as magnon \cite{katsura2010theory,onose2010observation,matsumoto2011theoretical}, phonon \cite{strohm2005phenomenological,sheng2006theory,kagan2008anomalous,zhang2010topological}, and magnetoelastic excitations \cite{park2019topological,zhang2019thermal}.
This so-called thermal Hall effect can be viewed as the thermal counterpart of the anomalous Hall effect.
The spin Hall effect also has a thermal analogy, known as the spin Nernst effect, in which magnons \cite{cheng2016spin,zyuzin2016magnon} or magnetoelastic excitations \cite{park2019thermal,zhang20193} transport spins in the direction transverse to the thermal gradient. 

In this work, we add a new item to the list of Hall effects, which we call the phonon angular momentum Hall effect (PAMHE), wherein the phonon angular momentum (PAM) flows transversely to the temperature gradient.
We show that the PAMHE is a ubiquitous phenomenon requiring only the presence of longitudinal and transverse phonon modes.
By introducing edges to the system, PAM accumulates at the edges as a result of the PAMHE.
When the atoms carry nonzero Born effective charge, the orbital motion of atoms creates magnetic moment.
Thus, edge PAM also induces edge magnetization, which can be measured by optical measurements.

\begin{figure*}[t]
\centering
\includegraphics[width=15cm]{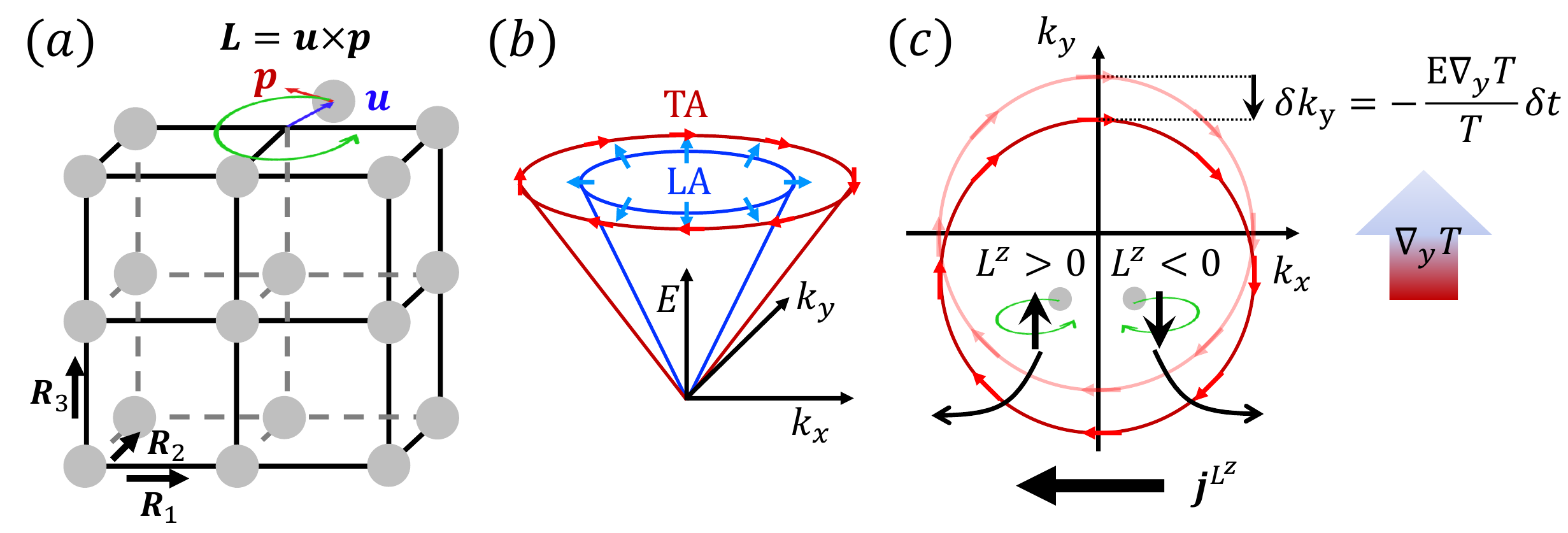}
\caption{\textbf{Illustration of phonon angular momentum Hall effect (PAMHE).} (a) Illustration of the phonon angular momentum (PAM) in a cubic lattice. 
(b) Transverse and acoustic phonon energy spectra and their polarization vectors, which are indicated by arrows, in a two-dimensional continuum.
(c) Shift of the transverse acoustic phonon states at a fixed energy $E$ by $\delta \bm{k}$ under the application of thermal gradient, and their subsequent dynamics.
}
\label{fig.PAM}
\end{figure*}

\textbf{Definition of phonon angular momentum (PAM)}.
To understand the PAM, it is useful to note that an atom in a lattice has three independent vibrational directions along the $\hat{\bm{x}}$, $\hat{\bm{y}}$, and $\hat{\bm{z}}$ axes, which can be likened to the $p_x$, $p_y$, and $p_z$ orbitals of an electron.
Just as these orbitals can be linearly combined to obtain states with nonzero orbital angular momentum, the vibrational modes of an atom can be linearly combined to obtain circularly polarized states with nonzero angular momentum, as shown in  Fig.~\ref{fig.PAM} (a).
To define PAM, let $\bm{u}_{\alpha}(\bm{R})$ and $\bm{p}_{\alpha}(\bm{R})$ be the displacement and momentum of an atom whose equilibrium position is given by $\bm{R}+\bm{\delta}_{\alpha}$, where $\bm{R}$ denotes the position of the unit cell and $\alpha$ is the sublattice index.
For convenience, we rescale the displacement (momentum) by multiplying (dividing) it by the square root of its mass.
The PAM is generally defined as \cite{zhang2014angular,juraschek2017dynamical,hamada2018phonon,juraschek2019orbital} $\bm{\mathcal{L}}=\sum_{\bm{R},\alpha} \bm{u}_{\alpha} (\bm{R}) \times \bm{p}_{\alpha}(\bm{R}).$
Defining $\bm{x}_\alpha(\bm{R})=(\bm{p}_{\alpha}(\bm{R}),\bm{u}_{\alpha}(\bm{R}))$  and taking the Fourier transformation $\bm{x}_{\bm{k}\alpha}=\frac{1}{\sqrt{V}}\sum_{\bm{R}} \bm{x}_{\alpha}(\bm{R})e^{-i\bm{k}\cdot (\bm{R}+\bm{\delta}_\alpha)}$, we have
\begin{equation}
\bm{\mathcal{L}}=\frac{1}{2}\sum_{\bm{k},\alpha} \bm{x}_{-\bm{k}\alpha}\bm{L}_\alpha \bm{x}_{\bm{k}\alpha}, \quad (L^\rho)_\alpha=\begin{pmatrix}
 & -\ell_{\rho}\\
 \ell_{\rho} & 
\end{pmatrix}. \label{eq.am_def}
\end{equation}
Here, $\ell_\rho$ is a matrix whose components are $(\ell_\rho)_{\mu\nu}=\varepsilon_{\mu\nu\rho}$, where $\varepsilon_{\mu \nu \rho} $ is the Levi-Civita symbol  and $\mu, \nu, \rho$ run over $x$, $y$, and $z$. 
Note that the matrix $(L^\rho)_\alpha$ is independent of $\alpha$.

\textbf{Generality of phonon angular momentum Hall effect (PAMHE)}.
To understand why PAMHE generally occurs, we consider the dynamics of phonon modes when thermal gradient is applied as in Ref.~\cite{go2018intrinsic}, where the dynamics of electronic orbitals under electric field is considered.
To gain the intuition, it suffices to restrict the discussion to a two-dimensional elastic continuum, so that the phonon Hamiltonian is given by 
\begin{equation}
\mathcal{H}=\frac{1}{2}\sum_{\bm{k}} \bm{x}_{-\bm{k}} H_{\bm{k}} \bm{x}_{\bm{k}}, \quad H_{\bm{k}}= \begin{pmatrix}
\tau_0 & 0 \\
0 & D_{\bm{k}}
\end{pmatrix}, \label{eq.phonon_H}
\end{equation}
where $\bm{x}_{\bm{k}}=\begin{pmatrix}
\bm{p}_{\bm{k}}, & \bm{u}_{\bm{k}}
\end{pmatrix}$, $\tau_0$ is a $2\times2$ identity matrix defined in the space spanned by the $x$ and $y$ components of $\bm{p}_{\bm{k}}$ or $\bm{u}_{\bm{k}}$, and $D_{\bm{k}}^{\mu \nu}=v_T^2 k^2 \delta_{\mu \nu}+(v_L^2-v_T^2)k_\mu k_\nu$  with $\mu, \nu=x,y$ is the dynamical matrix. 
The Heisenberg's equation is $i\frac{\partial}{\partial t} \bm{x}_{\bm{k}}(t)=(\tau_0\otimes\sigma^y) H_{\bm{k}} \bm{x}_{\bm{k}}(t)$, where $\sigma^i$ with $i=x,y,z$ are the Pauli matrices connecting $\bm{p}_{\bm{k}}$ and $\bm{u}_{\bm{k}}$.
In the following, we omit $\tau_0$ by writing $\sigma^i$ instead of $\tau_0\otimes\sigma^i$.
This equation has four normal modes $\bm{x}_{\bm{k},n}(t)=\bm{\chi}_{\bm{k},n} e^{-i E_{\bm{k},n} t}$, where $\bm{\chi}_{\bm{k},n}=\left( \begin{smallmatrix} -i E_{\bm{k},n} \bm{\epsilon}_{\bm{k},n}\\ \bm{\epsilon}_{\bm{k}, n}\end{smallmatrix} \right)$ for $n=L,T,-L,-T$, and the polarization vectors $\bm{\epsilon}_{\bm{k},n}$ satisfy $D_{\bm{k}}\bm{\epsilon}_{\bm{k},n}=E_{\bm{k},n}^2\bm{\epsilon}_{\bm{k},n}$.
Explicitly, $\bm{\epsilon}_{\bm{k},L}=\bm{\epsilon}_{\bm{k},-L}=\frac{1}{k\sqrt{2E_{\bm{k},L}}} \left( \begin{smallmatrix} k_x \\ k_y \end{smallmatrix} \right)$ and  
$\bm{\epsilon}_{\bm{k},T}=\bm{\epsilon}_{\bm{k},-T}=\frac{1}{k\sqrt{2E_{\bm{k},T}}}\left( \begin{smallmatrix} k_y \\ -k_x \end{smallmatrix} \right)$ are longitudinal and transverse polarization vectors, respectively, with $E_{\bm{k},L}=-E_{\bm{k},-L}=v_L k$ and $E_{\bm{k},T}=-E_{\bm{k},-T}=v_T k 
$, as shown in Fig.~\ref{fig.PAM} (b).  
Note that only two modes ($n=L,T$) are physically independent, but it is more convenient to utilize all four modes.
We also note that the polarization vectors are normalized such that $\bm{\chi}_{\bm{k},m}^\dagger \sigma^y \bm{\chi}_{\bm{k},n}=\delta^z_{m,n} $, where $\delta^z_{n,n}=1$ ($-1$) for $n=L,T$ ($n=-L,-T$), while $\delta^z_{m,n}=0$ for $m\neq n$.

Because we restrict ourselves to two dimensions, only $L^z$ is meaningful, which is now a $4\times 4$ matrix.
For the normal modes, the expectation value of PAM is $\langle \mathcal{L}^z \rangle_{\bm{k},n} (t) = \bm{x}^\dagger_{\bm{k},n} (t) L^z \bm{x}_{\bm{k},n} (t)=0$.
However, phonon states can develop nonzero angular momentum when thermal gradient is applied, as we now explain.
We first note that according to the method of pseudogravitational potential, the effect of thermal gradient can be treated by introducing a scalar potential $\phi(\bm{r})$ that couples to the energy $E$ in the form $\phi E$, with $\bm{\nabla}\phi=\frac{\bm{\nabla}T}{T}$ \cite{luttinger1964theory,matsumoto2011theoretical,matsumoto2014thermal,li2019intrinsic}.
Thus, under the application of thermal gradient $(\nabla_y T)\hat{\bm{y}}$ with $\nabla_y T>0$ for time $\delta t$, states shift in the momentum space from $\bm{k}$ to $\bm{k}+\delta \bm{k}$, where $\delta \bm{k}=(0,-\delta k_y)$ ($\delta k_y>0$) and $\delta k_y= E\bm{\nabla}\phi \delta t=\frac{E \nabla_y T}{T} \delta t$, as illustrated in Fig.~\ref{fig.PAM} (c). 

Focusing on the transverse modes, we note that, although $\bm{\chi}_{\bm{k},T}$ is not an eigenmode after the momentum shift $\bm{k} \rightarrow \bm{k}+\delta \bm{k}$, it can be decomposed in terms of the eigenmodes at $\bm{k}+\delta \bm{k}$, as $\bm{\chi}_{\bm{k},T}=\sum_{n}\alpha^T_{n} \bm{\chi}_{\bm{k}+\delta\bm{k},n}$, where $\alpha^T_{n}=\delta^z_{n,n}\bm{\chi}^\dagger_{\bm{k}+\delta\bm{k},n} \sigma^y \bm{\chi}_{\bm{k},T}$.
Then, the evolution of $\bm{\chi}_{\bm{k},T}$ for time $t$ gives $\bm{x}_{\bm{k},T}^{\textrm{shift}}(t)=\sum_{n}e^{-it E_{\bm{k}+\delta \bm{k},n}} \alpha^T_{n} \bm{\chi}_{\bm{k}+\delta\bm{k},n}$.
Even though the normal modes have vanishing expectation value of PAM,  the shifted states develop nonzero PAM $\langle \mathcal{L}^z\rangle_{\bm{k},T}^{\textrm{shift}} (t) = (\bm{x}_{\bm{k},T}^{\textrm{shift}})^\dagger(t) L^z \bm{x}_{\bm{k},T}^{\textrm{shift}}(t) \propto -k_x \delta k_y t$, to the lowest order in $t$ and $\delta k_y$ (see Methods).
Thus, the states with $k_x<0$, which drift towards $-\hat{\bm{x}}$, have positive expectation value of $\mathcal{L}^z$, while the states with  $k_x>0$, which drift towards $\hat{\bm{x}}$, have negative expectation value of $\mathcal{L}^z$, so that PAM flows towards $-\hat{\bm{x}}$.

For the case of longitudinal modes, a similar analysis shows that the PAM current is opposite in direction, i.e. towards $\hat{\bm{x}}$. 
However, the longitudinal modes generally have higher energy, so that they will have lower occupation.
Thus, the PAM current from the longitudinal modes only partially cancels that from the transverse modes, so that the net current direction is towards $-\hat{\bm{x}}$.

\begin{figure*}[t]
\centering
\includegraphics[width=15cm]{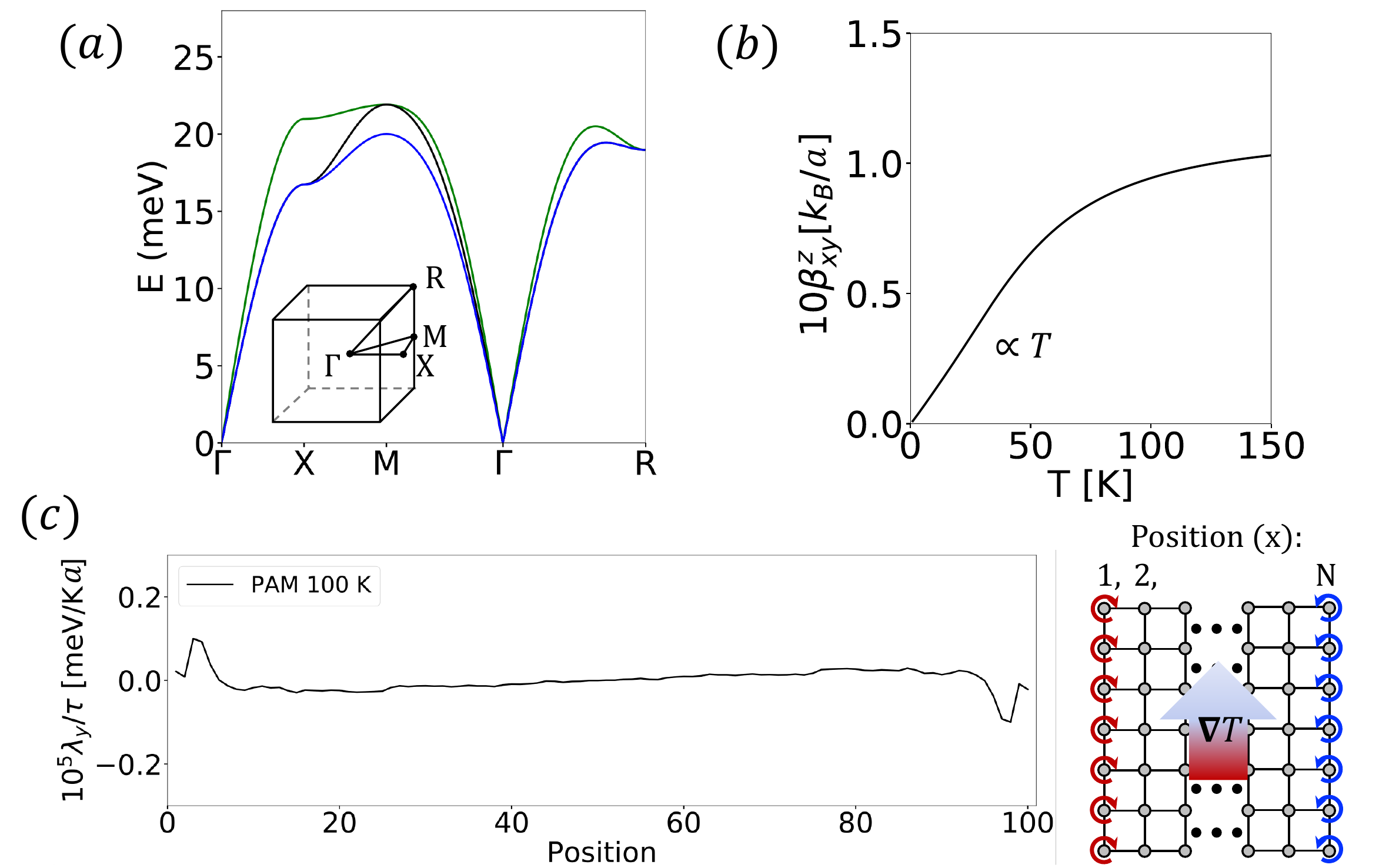}
\caption{\textbf{PAMHE in cubic lattice.} (a) Phonon spectrum along the high symmetry lines for the simple cubic lattice, with the NN transverse and longitudinal spring constants ($k_T=20~(\textrm{meV})^2$, $k_L=50~(\textrm{meV})^2$) and the NNN transverse and longitudinal spring constants ($k_T'=10~(\textrm{meV})^2$, $k_L'=20~(\textrm{meV})^2$). 
(b) The PAM Hall conductivity with the parameters given in (a) for the simple cubic lattice.
In the $y$ axis, $a$ is the lattice constant.
(c) Spatial distribution of PAM induced by temperature gradient for the simple cubic lattice with the geometry given on the right, which shows the two-dimensional projection of the simple cubic lattice onto the xy plane.
}
\label{fig.cube}
\end{figure*}

The above result can straightforwardly be generalized to three-dimensional continuum, which yields essentially the same conclusion that PAM $\langle \mathcal{L}^z \rangle$ flows towards $-\hat{\bm{x}}$ (see Methods).
Actually, for the continuum model, it is possible to give a simple argument for the result that PAM $\langle \mathcal{L}^z \rangle$ with opposite signs should flow in opposite directions by utilizing the presence of the mirror symmetry $\mathcal{M}_x$ about the plane normal to the $x$ axis.
Since $\mathcal{M}_x$ is not broken by the temperature gradient in the $y$ direction, it forces both the induced PAM $\langle \mathcal{L}^z \rangle$ and the velocity along the $x$ axis to have opposite signs at $k_x$ and $-k_x$.

\textbf{Linear response theory in cubic lattice}. 
Let us quantitatively evaluate the PAMHE  for a simple harmonic oscillator model on the simple cubic lattice in Fig.~\ref{fig.PAM} (a). 
For simplicity, we only consider the longitudinal and transverse spring constants between the nearest neighbors (NN) and the next nearest neighbors (NNN) to obtain the dynamical matrix $D_{\bm{k}}$ in the Hamiltonian (see Eq.~(\ref{eq.phonon_H})), whose explicit form is given in Supplementary Information (SI) \cite{supplement}.
Because we are working in three dimensions, there are six eigenstates $\bm{\chi}_{\bm{k},n}$ ($n=3,2,1,-1,-2,-3$), which we denote as $|n,\bm{k}\rangle$.
The eigenstates satisfy $\sigma^y H_{\bm{k}} |n,\bm{k}\rangle=E_{\bm{k},n}|n,\bm{k}\rangle$ and  $\langle n,\bm{k}| \sigma^y |n,\bm{k}\rangle =\delta^z_{m,n}$, where the eigenvalues satisfy $E_{\bm{k},n} \geq 0$ for $n>0$ and $E_{\bm{k},-n}=-E_{\bm{k},n}$.
The energy spectrum (eigenvalues with $n>0$) is shown in Fig.~\ref{fig.cube} (a). 
We note that the expectation value of the PAM for these states is $\langle n,\bm{k}| L^z | n,\bm{k}\rangle=0$ because of the time reversal and inversion symmetries, (see Methods). 
However, as we argued previously, application of temperature gradient dynamically induces PAM current, which we  evaluate below.

To study the transport of PAM, we define the PAM current density in the $\mu$ direction by $j_\mu^{L^\rho}=\frac{1}{2V} \sum_{\bm{k}} \bm{x}_{-\bm{k}} \frac{L^\rho \sigma^{y} v_{\bm{k},\mu}+v_{\bm{k},\mu} \sigma^{y} L^\rho }{2} \bm{x}_{\bm{k}}$, where $V$ is the volume.
We note that this definition of PAM current corresponds to the conventional definition of spin current in the context of spin Hall effect~\cite{sinova2015spin}, and similarly, it does not satisfy the continuity equation because PAM is not a conserved quantity, i.e.  $[\mathcal{L},\mathcal{H}] \neq 0$~\cite{li2019intrinsic,shi2006proper} (see Methods).
To compute the PAM Hall conductivity, we notice that the phonon Hamiltonian $\mathcal{H}$ is a bosonic BdG Hamiltonian as explained in Methods, so that we  can directly apply the linear response theory in Ref.~\cite{li2019intrinsic}, which is reviewed in the SI \cite{supplement}:
the expectation value of the PAM current to the linear order in the thermal gradient $\nabla_\nu T$ is $\langle j_{\mu}^{L^\rho}\rangle_{\textrm{neq}}=-\beta^{\rho}_{\mu \nu}\nabla_{\nu} T$, in which
\begin{equation}
\beta^{\rho}_{\mu \nu}=\frac{k_B }{V}\sum_{\bm{k}}\sum_{n>0} \Omega_{\mu \nu,n}^\rho (\bm{k})c_1(E_{\bm{k},n}), \label{eq.PAMH_conductivity}
\end{equation}
where we have defined the PAM curvature
$\Omega^{\rho}_{\mu \nu,n} (\bm{k})= \hbar\sum_{m}' \frac{\delta^z_{m,m} \textrm{Im}[\langle n,\bm{k}| L^\rho \sigma^y v_{\bm{k},\mu}+v_{\bm{k},\mu} \sigma^y L^\rho | m, \bm{k} \rangle \langle m, \bm{k}| v_{\bm{k},\nu}| n,\bm{k} \rangle ]}{(E_{\bm{k},n}-E_{\bm{k},m})^2} $ and the notation$~'$ is used to indicate that the sum excludes $m=n$. Also, $v_{\bm{k},\mu}=\frac{1}{\hbar} \frac{\partial H_{\bm{k}}}{\partial k_\mu}$, $c_1(x)=(1+g(x)) \log(1+g(x))-g(x)\log g(x)$, and $g(x)=\frac{1}{e^{x/k_BT}-1}$ is the Bose-Einstein distribution.
We show $\beta_{xy}^z$ calculated for the simple cubic lattice model in Fig.~\ref{fig.cube} (b).
We see that under the temperature gradient $(\nabla_y T)\hat{\bm{y}}$ with $\nabla_y T>0$, PAM flows in the $-\hat{\bm{x}}$ direction, as heuristically argued previously. 
We also note that at $T=100$K, $\beta_{xy}^z \approx 0.1 [k_B/a]$ (see Fig.~\ref{fig.cube} (b)) is about $10^3$ times larger than the magnon spin Nernst coefficient in Ref.~\cite{cheng2016spin} obtained for a hexagonal antiferromagnet.

We also notice that $\beta_{xy}^z \propto T$ at low temperature.
This is because $\Omega^{z}_{xy,n}(\bm{k} )\sim \frac{1}{k^2}$ \cite{supplement}, so that $\beta^{z}_{xy}$ scales with temperature as $\int k^{d-3}dk c_1[(g(\frac{vk}{k_BT}))] \sim T^{d-2}$, where $d\geq 2$ is the spatial dimension and $v$ is a constant (phonon velocity). Thus, $\beta^z_{xy}\propto T$ for $d=3$.
On the other hand, the integral diverges logarithmically for $d=2$ (see Methods).

From the diffusion theory~\cite{murakami2003dissipationless}, we can expect accumulation of PAM at the edges proportional to the PAM lifetime and the PAM Hall current.
However, because PAM is not conserved, there may not be a simple correspondence between PAMHE and edge PAM accumulation \cite{shi2006proper}, so that we should directly compute the nontrivial PAM accumulation by introducing edges, as shown in Fig.~\ref{fig.cube} (c).
Let $L_x^z$ be the $z$-component of the PAM density of the atoms lying in the $yz$ plane passing through the position $x$ (see Fig.~\ref{fig.cube} (c)).
The Boltzmann transport theory with constant relaxation time $\tau$ gives \cite{ashcroft1976solid,mook2019spin} $g_{\textrm{neq}}(E)= g_{\textrm{eq}}(E)-\tau \bm{v}\cdot \bm{\nabla}_\nu T \frac{E}{k_BT^2}\frac{e^{E/k_BT}}{(e^{E/k_BT}-1)^2}$.
In using the Boltzmann transport theory, we assume that phonon quickly relaxes to local equilibrium via anharmonicity in the crystal. 
It is also sufficient to use constant relaxation time approximation for a rough estimation, and the dependence of the relaxation time on the phonon branches and on crystal momentum does not alter the conclusion that PAM accumulates at the edges.
Thus,  the PAM density induced by temperature gradient $(\nabla_y T) \hat{\bm{y}}$ is $\langle L^z_x\rangle_{\textrm{neq}}-\langle L^z_x \rangle_{\textrm{eq}}=-\lambda_{y}(x)\nabla_{y} T$, where
\begin{align}\label{eq.lambda}
\lambda_{y}(x)=&\frac{\tau}{2k_BT^2}\frac{1}{V} \sum_{\bm{k}}\sum_{n=-N}^{N} \langle n,\bm{k}| L_x^z | n,\bm{k} \rangle \times\nonumber \\ 
&\langle n,\bm{k}| v_{\bm{k},y}|n,\bm{k}\rangle\frac{ \delta^z_{n,n} E_{\bm{k},n} e^{\delta^z_{nn} E_{\bm{k},n}/k_BT}}{(e^{\delta^z_{nn} E_{\bm{k},n}/k_BT}-1)^2}.
\end{align}
The spatial distribution of the PAM is shown in Fig.~\ref{fig.cube} (c).

Although the PAM is difficult to observe directly, if the atoms have nonzero Born effective charge $Z_{\textrm{eff}}$, the PAM will generate phonon magnetic moment (PMM) given by $\frac{eZ_{\textrm{eff}}}{2M}L^z$, where $e$ is the elementary charge and $M$ is the mass of the atom.
This is an important consequence because magnetic moment can be directly observed through optical means such as the magneto-optical Kerr microscopy \cite{kato2004observation,stamm2017magneto}.
Because the simple cubic lattice has only one atom per unit cell, we cannot expect the PAM to generate magnetization.
We therefore consider next the CsCl lattice structure with two atoms per unit cell.
\begin{figure}[t]
\centering
\includegraphics[width=8.5cm]{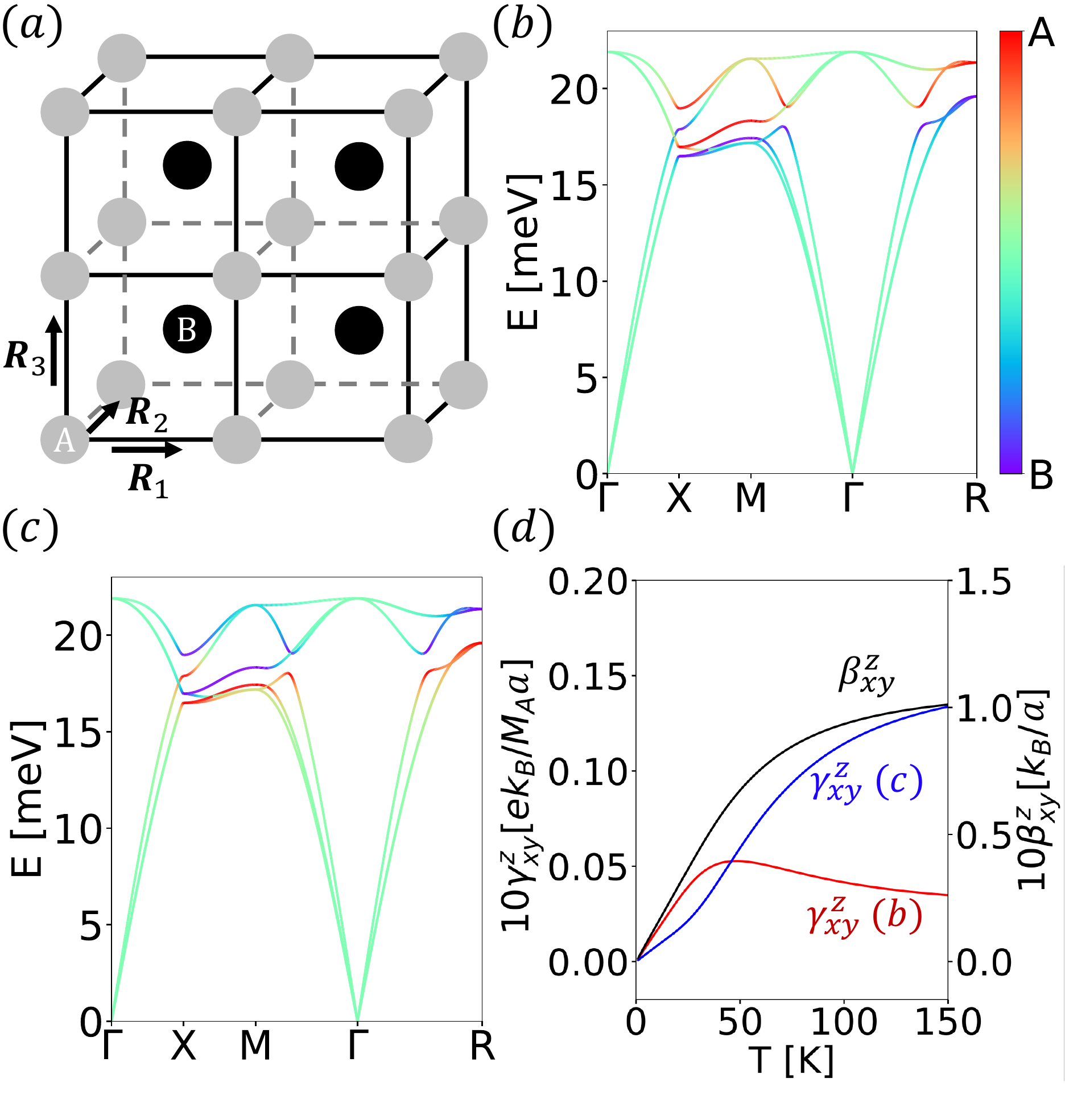}
\caption{\textbf{PAMHE and PMMHE in CsCl lattice.} (a) The CsCl lattice, where the grey (black) circles indicate $A$ ($B$) sites. 
(b) The energy spectrum with the NN transverse and longitudinal spring constants ($k_T=20~(\textrm{meV})^2$ and $k_L=50~(\textrm{meV})^2$) and the NNN transverse and longitudinal spring constants between the $A$ sites ($k^{A}_{T}=12~(\textrm{meV})^2$ and $k^{A}_{L}=30~(\textrm{meV})^2$) and $B$ sites ($k^{B}_{T}=8~(\textrm{meV})^2$ and $k^{B}_{L}=20~(\textrm{meV})^2$).
The color represents the wavefunction content of $A$ and $B$.
(c) Energy spectrum and wavefunction content of $A$ and $B$ with the same parameters used for (b) except that the spring constant values between $A$ and $B$ sites are interchanged, i.e. $k^{B}_{T}=12~(\textrm{meV})^2$, $k^{B}_{L}=30~(\textrm{meV})^2$, $k^{A}_{T}=8~(\textrm{meV})^2$, and $k^{A}_{L}=20~(\textrm{meV})^2$.
(d) The PMM Hall conductivities $\gamma^z_{xy}$ for the parameters in (b) (red curve)  and (c) (blue curve) and the PAM Hall conductivity $\beta^z_{xy}$ (black curve), which is equivalent for the two sets of parameters.
}
\label{fig.cscl}
\end{figure}

\textbf{Results for CsCl lattice}.
Consider the CsCl lattice structure in Fig.~\ref{fig.cscl} (a) with two interpenetrating simple cubic lattices formed by atoms $A$ and $B$.
Let us assume that the atomic mass of $A$ ($B$) is $M_A$ ( $M_B=1.5M_A$) and that the Born effective charge of $A$ ($B$) is $Z_{\textrm{eff}}^{A}=1$ ($Z_{\textrm{eff}}^{B}=-1$). 
Thus, the magnetic moments of the atoms are given by   $\mu_\alpha L^z_\alpha$ with $\mu_\alpha=\frac{Z_{\textrm{eff}}^{\alpha}e}{2M_\alpha}$ for $\alpha=A,B$.
As before, we consider only the longitudinal and transverse spring constants between the NN and the NNN (see SI \cite{supplement}).
The energy spectrum is shown in Fig.~\ref{fig.cscl} (b), where the values of the spring constants are given in the caption.
The corresponding PAM Hall conductivity $\beta^z_{xy}$ is shown in Fig.~\ref{fig.cscl} (d) with black curve, which shows a similar behavior to Fig.~\ref{fig.cube} (b) calculated for the simple cubic lattice.

Because $\mu_{A,B}\neq 0$ for the atoms in the CsCl lattice, we can define the PMM Hall effect (PMMHE) in addition to the PAMHE, wherein the temperature gradient causes a transverse flow of PMM. 
The discussion for PMM is parallel to that for PAM.
 Let $m^\rho$ be the matrix of magnetic moments with $(m^\rho)_\alpha=\mu_\alpha (L^\rho)_\alpha$ (see Eq.~\eqref{eq.am_def}).
We define the PMM current as $j_\mu^{m^\rho}=\frac{1}{2V} \sum_{\bm{k}} \bm{x}_{-\bm{k}} \frac{m^\rho \sigma^{y} v_{\bm{k},\mu}+v_{\bm{k},\mu} \sigma^{y} m^\rho }{2} \bm{x}_{\bm{k}}$.
The PMM conductivity $\gamma^\rho_{\mu \nu}$ is given by the expression $\langle j_{\mu}^{m^\rho}\rangle_{\textrm{neq}}=-\gamma^{\rho}_{\mu \nu}\nabla_{\nu} T$, where
\begin{equation}
\gamma^{\rho}_{\mu \nu}=\frac{k_B }{V}\sum_{\bm{k}}\sum_{n>0} \Lambda_{\mu \nu,n}^\rho (\bm{k})c_1(E_{\bm{k},n}), \label{eq.PMMH_conductivity}
\end{equation}
and we have defined the PMM curvature
$\Lambda^{\rho}_{\mu \nu,n} (\bm{k})= \hbar\sum_{m}' \frac{\delta^z_{m,m} \textrm{Im}[\langle n,\bm{k}| m^\rho \sigma^y v_{\bm{k},\mu}+v_{\bm{k},\mu} \sigma^y m^\rho | m, \bm{k} \rangle \langle m, \bm{k}| v_{\bm{k},\nu}| n,\bm{k} \rangle ]}{(E_{\bm{k},n}-E_{\bm{k},m})^2}$.
We see that $\gamma^z_{xy}$ ( Fig.~\ref{fig.cscl} (d) (red curve)) shows a very different behavior from $\beta^z_{xy}$.

\begin{figure*}[t]
\centering
\includegraphics[width=15cm]{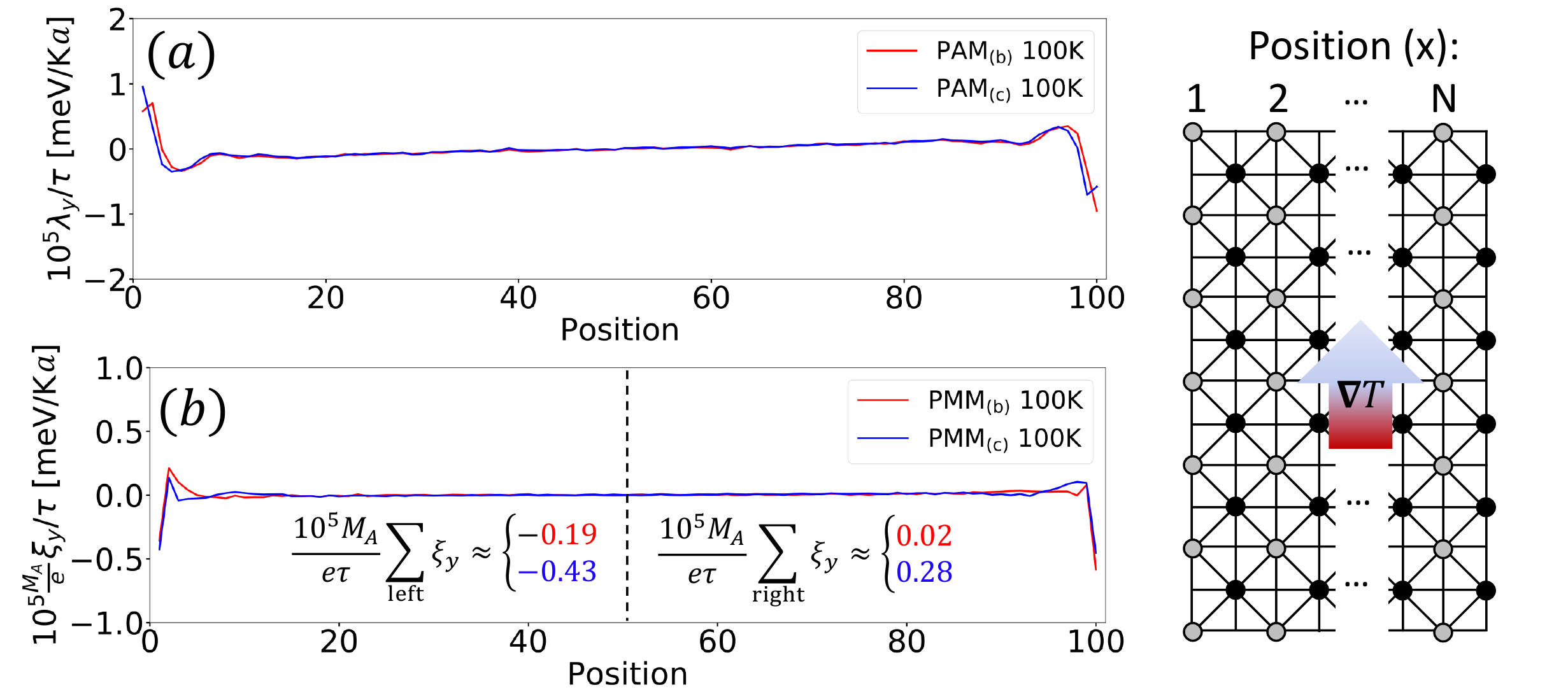}
\caption{\textbf{PAM and PMM accumulation in CsCl lattice.} (a) PAM accumulation with the configuration shown on the right, which shows the projection of the CsCl lattice onto the xy plane. The red and the blue lines are calculated with the parameters in Fig.~\ref{fig.cscl} (b) and (c) respectively. 
(b) PMM accumulation, where the red and the blue lines are calculated with the parameters in Fig.~\ref{fig.cscl} (b) and (c) respectively. 
The sum of $\frac{10^5M_A}{e\tau}\xi_y$  for the left half of the system is $-0.19~\textrm{meV/K}a$ ($-0.43~\textrm{meV/K}a$) for the red (blue) curve, and the sum for the right half of the system is $0.02~\textrm{meV/K}a$ ($0.28~\textrm{meV/K}a$) for the red (blue) curve.
}
\label{fig.cscl_acc}
\end{figure*}

To understand the behavior of $\gamma^z_{xy}$, we notice that the spring constants between the $A$ sites are stiffer than between the $B$ sites, $\mu_A\cdot\mu_B<0$, and $\mu_A > |\mu_B|$.
In addition, for energy less than approximately $7$ meV, the phonon wavefunction is nearly equally shared between the $A$ and the $B$ sites, as shown in Fig.~\ref{fig.cscl} (b).
However, the PAM arising from $A$ sites is expected to be transported faster than PAM arising from $B$ sites because the spring constants between $A$ sites are stiffer than those between $B$ sites.
Thus, $\gamma^z_{xy}$ shown in Fig.~\ref{fig.cscl} (d) (red curve) has the same sign as $\beta^z_{xy}$ at low temperature because the contribution to $\gamma^z_{xy}$ from the $A$ sites with positive Born effective charge is larger.
However, at higher energy, $B$ sites have significantly higher occupation so that $\gamma^{z}_{xy}$ decreases with increasing temperature, since the contribution to $\gamma^{z}_{xy}$ from $B$ sites with negative Born effective charge significantly increases.

To gain a deeper understanding of the behavior of $\gamma^z_{xy}$, it is convenient to interchange the values of the NNN spring constants between the $A$ sites and those between the $B$ sites.
Under this operation, the energy spectrum and $\beta^z_{xy}$ remain the same as before, while the  wavefunction contents of $A$ and $B$ sites switch, as shown in Fig.~\ref{fig.cscl} (c) \cite{supplement}.
Unlike $\beta^{z}_{xy}$, the behavior of $\gamma^{z}_{xy}$ changes significantly, as can be seen in Fig.~\ref{fig.cscl} (d) (blue curve).
To explain this, note that for energy approximately less than $7$ meV, the wavefunction is nearly equally shared between $A$ and $B$, while the spring constants between the $B$ sites are stiffer than those between $A$ sites.
Therefore, the PAM arising from $B$ sites are transported faster, so that $\gamma^{z}_{xy}$ is smaller than that obtained previously at low temperature, although we still have $\gamma^{z}_{xy}>0$ because $\mu_A>|\mu_B|$.
At higher energy, the wavefunction content is higher for $A$ sites, so that the value of $\gamma^{z}_{xy}$ keeps increasing with temperature.

To illustrate the implications of the PAM and PMM Hall currents, we compute the spatial distribution of PAM $\lambda_{y}(x)$ in Eq.~\eqref{eq.lambda}, and that of PMM $\xi_{y}(x)$, which is defined from the expression $\langle m^z_x\rangle_{\textrm{neq}}-\langle m^z_x \rangle_{\textrm{eq}}=-\xi_{y}(x)\nabla_{y} T$.
Using the Boltzmann transport theory as in Eq.~\eqref{eq.lambda}, we have
\begin{align}\label{eq.xi}
\xi_{y}(x)=&\frac{\tau}{2k_BT^2}\frac{1}{V} \sum_{\bm{k}}\sum_{n=-N}^{N} \langle n,\bm{k}| m_x^z | n,\bm{k} \rangle \times\nonumber \\ 
&\langle n,\bm{k}| v_{\bm{k},y}|n,\bm{k}\rangle\frac{ \delta^z_{n,n} E_{\bm{k},n} e^{\delta^z_{nn} E_{\bm{k},n}/k_BT}}{(e^{\delta^z_{nn} E_{\bm{k},n}/k_BT}-1)^2}.
\end{align}
Note that two planes of atoms, each consisting of either $A$ or $B$, are contained in a single position index, as shown in Fig.~\ref{fig.cscl_acc}.
We show $\lambda_y(x)$ and $\xi_{y}(x)$ in Fig.~\ref{fig.cscl_acc} (a) and (b), respectively, computed using the two sets of parameters discussed above, which were used in Fig.~\ref{fig.cscl} (b) and (c). 
In Fig.~\ref{fig.cscl_acc} (a), we see that the spatial distribution of PAM does not differ much between the two sets of parameters, which is not surprising as the two cases show the same PAM Hall conductivity.
However, the different behaviors of PMM Hall conductivity $\gamma^z_{xy}$ in Fig.~\ref{fig.cscl} (d) for the two sets of parameters foretell different behaviors for the distribution of PMM, as shown in Fig.~\ref{fig.cscl_acc} (b). 

We find that the difference in the PMM accumulation at the left and the right edges for the red curve in Fig.~\ref{fig.cscl_acc} (b) is significantly smaller than that for the  blue curve, which agrees with the behavior of $\gamma^z_{xy}$ (red and blue curves in Fig.~\ref{fig.cscl} (d)).
We further note that the induced PMM near the edges has order of magnitude of $10^{-2}\tau/(1s)$ Bohr magneton per unit cell for the blue curve, where we have assumed that the unit cell size is $5\textrm{\AA}$, $M_A$ is $25$ amu, and the temperature gradient is $10$K/$100\mu$m.
If we assume that the acoustic phonon lifetime is $10\sim 100$ ps ~\cite{togo2015distributions}, the PMM accumulation can be expected to be around $10^{-13}\sim10^{-12}$ Bohr magneton per unit cell.

\textbf{Discussion.}
Up to now, we have only considered the PAM Hall conductivity $\beta_{xy}^{\rho}$ for $\rho= z$.
This is because the components with $\rho =x,y$ vanish for the simple cubic lattice due to lattice symmetries.
In particular, the mirror symmetry $\mathcal{M}_z:z\rightarrow -z$ or $\mathcal{M}_x:x\rightarrow -x$ forbids $\beta^x_{xy}$, while $\mathcal{M}_z$ or $\mathcal{M}_y:y\rightarrow -y$ forbids  $\beta^y_{xy}$.
However, $\beta^x_{xy}$ and $\beta^y_{xy}$ are not forbidden by the time reversal symmetry, so that they do not vanish for systems with lower crystalline symmetry.

On the other hand, we have argued that $\beta_{xy}^{\rho}$ for $\rho= z$ relies only on the presence of the longitudinal and transverse phonon modes.
This implies not only that PAMHE is a general phenomenon, but also that $\beta_{xy}^{z}$ behaves similarly across various systems.
In contrast, the behavior of PMMHE is not universal, but we can expect it to be sizable in ionic crystals if the masses of the ions with opposite charge differ significantly. 

We also showed that the PAMHE and PMMHE induce edge PAM and PMM accumulation.
The PMM accumulation should be important especially in insulators because no edge magnetization is expected from electrons, even in spin~\cite{murakami2003spin} and orbital~\cite{canonico2019two,canonico2020orbital} Hall insulators~\cite{murakami2003spin}.
In magnetic insulators, the PMMHE can be a significant source of edge magnetization in addition to that from spin Nernst effect.
Since the spin-phonon interaction may significantly modify the PAM~\cite{zhang2014angular}, its role on the PAMHE would be an interesting topic for future study.
We believe that manipulating phonon angular momentum will open up an  avenue for developing functional materials based on phonon engineering and that PAMHE provides an important principle towards this goal.

\section{Methods}

\textbf{Second quantization and BdG formalism}. 
The phonon creation and annihilation operators are introduced by writing \begin{equation}
\bm{x}_{\bm{k}}=\sum_{n}\bm{\chi}_{\bm{k},n}b_{\bm{k},n},
\end{equation}
 where $n=N,..., 1, -1,..., -N$, and $N$ is the number of phonon modes with positive frequency ~\cite{park2019topological,zhang2010topological}. 
Here, $b_{\bm{k},n}$ is the phonon annihilation operator satisfying $b^\dagger_{\bm{k},n}=b_{-\bm{k},-n}$, and $\bm{\chi}_{\bm{k},n}$ is the phonon wavefunction satisfying 
$\sigma^y H_{\bm{k}} \bm{\chi}_{\bm{k},n}=E_{\bm{k},n}\bm{\chi}_{\bm{k},n}, \quad \bm{\chi}_{\bm{k},m}^\dagger \sigma^y \bm{\chi}_{\bm{k},n}=\delta^z_{m,n}$
and $\bm{\chi}_{\bm{k},n}^*=\bm{\chi}_{-\bm{k},-n}$ (recall our definition that $\delta^z_{m,n}=1 (-1)$ for $m=n>0$ ($m=n<0$) and $0$ otherwise).
The eigenvalues satisfy $E_{\bm{k},n} \geq 0$ for $n>0$,  $E_{\bm{k},n} \leq 0$ for $n<0$, and $E_{\bm{k},n}=-E_{-\bm{k},-n}$.
Thus, $\mathcal{H}=\frac{1}{2}\sum_{\bm{k},n}b_{\bm{k},n}^\dagger b_{\bm{k},n} \delta^z_{n,n}E_{\bm{k},n}=\sum_{\bm{k},n>0} b^\dagger_{\bm{k},n}b_{\bm{k},n} (E_{\bm{k},n}+\frac{1}{2})$.

From the field operator $\bm{x}_{\bm{k}}=\left(\begin{smallmatrix} \bm{p}_{\bm{k}} \\ \bm{u}_{\bm{k}}\end{smallmatrix}\right)$, we can obtain the bosonic BdG field operators $\bm{y}_{\bm{k}}$ by making the transformation $\bm{y}_{\bm{k}}=\frac{\sqrt{2}}{2}\left(\begin{smallmatrix}\bm{p}_{\bm{k}}-i\bm{u}_{\bm{k}}\\
\bm{p}_{\bm{k}}+i\bm{u}_{\bm{k}} \end{smallmatrix}\right) =U \bm{x}_{\bm{k}}$.
Then, $[y_{\bm{k},i}^\dagger,y_{\bm{k}',j}]=-\delta_{i,j}\delta_{\bm{k},\bm{k}'}$ and $y_{\bm{k},i}^\dagger=\sum_{j}\sigma^x_{ij}y_{-\bm{k},j}$ (here, $\delta_{i,j}=1$ if $i=j$ and $0$ otherwise).
Making the same transformation to the eigenvectors $\bm{\xi}_{\bm{k},n}=U \bm{\chi}_{\bm{k},n}$, we have $\sigma^z H_{\bm{k}} \bm{\xi}_{\bm{k},n}=E_{\bm{k},n}\bm{\xi}_{\bm{k},n}$ and $\bm{\xi}_{\bm{k},m}^\dagger \sigma^z \bm{\xi}_{\bm{k},n}=\delta^z_{m,n}$ \cite{park2019topological}.
It is convenient to know that this relation exists between the phonon and bosonic BdG Hamiltonian, as the expressions for the transport coefficients calculated for a general bosonic BdG Hamiltonian can directly be applied to phonon Hamiltonian as well.

\textbf{Properties of PAM $\bm{\mathcal{L}}$}.
Let us first explain why of the PAM is not conserved by computing $[\bm{\mathcal{L}},\mathcal{H}]$.
Noting that $[x_{\bm{k}m},x_{\bm{k}'n}]=-\sigma^y_{mn}\delta_{\bm{k},-\bm{k}'}$, $\bm{\mathcal{L}}=\frac{1}{2}\sum_{\bm{k}}\bm{x}_{-\bm{k}} \bm{L} \bm{x}_{\bm{k}}$, $\mathcal{H}=\frac{1}{2}\sum_{\bm{k}}\bm{x}_{-\bm{k}} H_{\bm{k}}\bm{x}_{\bm{k}}$, and $H_{\bm{k}}=H^T_{-\bm{k}}$, we find
\begin{align}
[\bm{\mathcal{L}},\mathcal{H}]=&\frac{1}{4}\sum_{\bm{k}} \bm{x}_{-\bm{k}}[H_{\bm{k}}i\sigma^y \bm{L}+H^T_{-\bm{k}}i\sigma^y \bm{L}\nonumber \\
&~~~~~~~~~~~~~~~~~~~-i\sigma^y\bm{L}H_{\bm{k}}-i\sigma^y\bm{L} H^T_{-\bm{k}}]\bm{x}_{\bm{k}} \nonumber \\
&=\frac{1}{2} \sum_{\bm{k}}\bm{x}_{-\bm{k}}[H_{\bm{k}},i\sigma^y \bm{L}]\bm{x}_{\bm{k}}.
\end{align}
Since $[H_{\bm{k}},i\sigma^y \bm{L}]\neq 0$ in general, which can easily be checked for all of the models we have used in this work, PAM is not conserved. 
Another way of stating this is that the energy eigenstates are not eigenstates of PAM, because this would require  $\sigma^y H_{\bm{k}} |n,\bm{k}\rangle=E_{\bm{k},n} |n,\bm{k}\rangle$ and $\sigma^y \bm{L} |n,\bm{k}\rangle=\bm{L}_{\bm{k},n}|n,\bm{k}\rangle$, clearly forbidden by the relation $[H_{\bm{k}},i\sigma^y \bm{L}]\neq 0$.

The conserved operator in the (point) mass and spring model is instead the total angular momentum $\bm{\mathcal{L}}^{\textrm{tot}}=\sum_{\bm{R},\alpha}\bm{r}_\alpha(\bm{R})\times \bm{\pi}_{\alpha}(\bm{R})$, where $\bm{r}_\alpha(\bm{R})=\bm{R}+\bm{\delta}_{\alpha} (\bm{R})+\bm{u}_{\alpha}(\bm{R})$ and $\bm{\pi}_{\alpha}(\bm{R})=M_\alpha \frac{d}{dt} \bm{r}_\alpha(\bm{R})=M_\alpha \frac{d}{dt}(\bm{R}+\bm{\delta}_\alpha (\bm{R}))+ \bm{p}_\alpha(\bm{R})$. Note that we have not rescaled $\bm{u}_{\alpha}(\bm{R})$ and $\bm{p}_{\alpha}(\bm{R})$ by the atomic mass $M_\alpha$ as was done in the Results. 
To understand the relation between the PAM and the total angular momentum, let us decompose the total angular momentum into the PAM ($\bm{\mathcal{L}}$) and the remaining terms ($\bm{\mathcal{L}}^{\textrm{lat}}$): $\bm{\mathcal{L}}^{\textrm{tot}}=\bm{\mathcal{L}}+\bm{\mathcal{L}}^{\textrm{lat}}$, where $\bm{\mathcal{L}}^{\textrm{lat}}=\sum_{\bm{R},\alpha}[(\bm{R}+\bm{\delta}_{\alpha} (\bm{R}))\times \bm{p}_{\alpha}(\bm{R})+\bm{u}_\alpha(\bm{R})\times M_\alpha \frac{d}{dt}(\bm{R}+\bm{\delta}_\alpha (\bm{R}))]+\sum_{\bm{R},\alpha} (\bm{R}+\bm{\delta}_\alpha (\bm{R}))\times M_\alpha \frac{d}{dt}(\bm{R}+\bm{\delta}_\alpha (\bm{R}))$.
To understand the physical meaning of $\bm{\mathcal{L}}^{\textrm{lat}}$, let us assume that the periodic motion of phonon is much faster than the motion of equilibrium positions. 
Averaging over timescale long compared to phonon period, we have $\langle \bm{\mathcal{L}}^{\textrm{lat}}\rangle_{\textrm{t}}=\sum_{\bm{R},\alpha} (\bm{R}+\bm{\delta}_\alpha (\bm{R}))\times M_\alpha \frac{d}{dt}(\bm{R}+\bm{\delta}_\alpha (\bm{R}))$, which is nothing but the angular momentum of the lattice as a whole in the absence of lattice vibrations.
Thus, on average, the total angular momentum is the sum of the PAM due to the internal vibrations and the angular momentum due to the lattice as a whole.
If we also include the electronic degrees of freedom, the conserved total angular momentum should also include the spin and orbital degrees of freedom.

Let us examine some additional properties of the PAM.
Here, we will adopt the notation $|n,\bm{k}\rangle=\bm{\chi}_{\bm{k},n}$.
Because the PAM is not conserved, i.e. $[\sigma^y \bm{L},H_{\bm{k}}]\neq 0$, we have $\langle m, \bm{k}| [\sigma^y \bm{L}, H_{\bm{k}}]|n, \bm{k}\rangle=\langle m, \bm{k}| \bm{L}|n, \bm{k}\rangle (E_{\bm{k},n} - E_{\bm{k},m}) \neq 0$ for some $m\neq n$.
Thus, 
\begin{equation}
\bm{\mathcal{L}}=\frac{1}{2} \sum_{\bm{k},m,n} \langle m, \bm{k}|\bm{L} |n,\bm{k}\rangle b_{\bm{k},m}^\dagger b_{\bm{k},n} \label{eq.PAM_SQ}
\end{equation}
has interband components in the energy basis.
Furthermore, both cubic and CsCl lattices have spatial inversion ($\mathcal{P}$) and time reversal ($\mathcal{T}$)  symmetries.
The actions of these symmetry operators on the energy eigenstates are given by $\mathcal{P} |n, \bm{k}\rangle=-|n,-\bm{k}\rangle$ and $\mathcal{T}|n, \bm{k}\rangle=-\sigma^z \mathcal{K}|n, -\bm{k}\rangle$, where $\mathcal{K}$ is the complex conjugation operator.
Thus, in the presence of $\mathcal{PT}$ symmetry, $|n,\bm{k}\rangle = \sigma^z\mathcal{K} |n,\bm{k}\rangle$. 
Since $\bm{L}\sigma^z=-\sigma^z \bm{L}$ and $\bm{L}^\dagger=\bm{L}$, we have $\langle n,\bm{k}| \bm{L} |n, \bm{k}\rangle =0$, i.e. there are no intraband components. 
Note that the expectation value of the PAM in thermal equilibrium, that is, $\langle \bm{\mathcal{L}} \rangle_{\textrm{eq}}$, vanishes when there is time reversal symmetry.

\textbf{Angular momentum of shifted states}.
Let us derive the relation $\langle \mathcal{L}^z\rangle_{\bm{k},T}^{\textrm{shift}} (t)  \propto -k_x \delta k_y t$ for two-dimensional and three-dimensional continuum models.
We begin with the two-dimensional continuum model, for which the equation of motion for the normal modes $\sigma^y H_{\bm{k}} \bm{\chi}_{\bm{k},n}=E_{\bm{k},n}\bm{\chi}_{\bm{k},n}$ yields
\begin{equation}
\bm{\chi}_{\bm{k},n}=\begin{pmatrix}
-iE_{\bm{k},n} \bm{\epsilon}_{\bm{k},n} \\
\bm{\epsilon}_{\bm{k},n}
\end{pmatrix},
\end{equation}
where
\begin{align}
E_{\bm{k},L}&=-E_{\bm{k},-L}=v_L k \label{eq.E_L}\\
E_{\bm{k},T}&=-E_{\bm{k},-T}=v_T k \label{eq.E_T} \\
\bm{\epsilon}_{\bm{k},L}&=\bm{\epsilon}_{\bm{k},-L}=\frac{1}{k\sqrt{2E_{\bm{k},L}}}  \begin{pmatrix}
k_x \\ k_y
\end{pmatrix} \label{eq.eps_L}\\
\bm{\epsilon}_{\bm{k},T}&=\bm{\epsilon}_{\bm{k},-T}=\frac{1}{k\sqrt{2E_{\bm{k},T}}}\begin{pmatrix}
k_y \\ -k_x
\end{pmatrix}. \label{eq.eps_T}
\end{align}
It can be checked that the normalization condition $\bm{\chi}_{\bm{k},m}^\dagger \sigma^y \bm{\chi}_{\bm{k},n}=\delta^z_{m,n}$ is satisfied.
Also, the expression $\mathcal{L}^z=\frac{1}{2}\sum_{\bm{k},m,n} \langle m, \bm{k}|L^z |n,\bm{k}\rangle b_{\bm{k},m}^\dagger b_{\bm{k},n}$ given in Eq.~\eqref{eq.PAM_SQ} shows that the expectation value of the angular momentum for the longitudinal and transverse modes are given respectively by $\langle L,\bm{k}|L^z|L,\bm{k}\rangle$ and $\langle T,\bm{k}|L^z|T,\bm{k}\rangle$, where the properties $b^\dagger_{\bm{k},n}=b_{-\bm{k},-n}$ and $|-n,-\bm{k}\rangle=\mathcal{K}|n,\bm{k}\rangle$ were used to eliminate the (redundant) negative-energy modes.

Below, we show that the transverse mode with momentum shift $\bm{k}\rightarrow \bm{k}'=\bm{k}+\delta \bm{k}$, $\bm{x}_{\bm{k},T}^{\textrm{shift}}(t)=\sum_{n}e^{-it E_{\bm{k}',n}} \alpha^T_{n} \bm{\chi}_{\bm{k}',n}$, develops angular momentum $\langle \mathcal{L}^z\rangle_{\bm{k},T}^{\textrm{shift}} (t) = (\bm{x}_{\bm{k},T}^{\textrm{shift}})^\dagger(t) L^z \bm{x}_{\bm{k},T}^{\textrm{shift}}(t) \propto -k_x \delta k_y t$, to linear order in $t$.
For this, it is useful to note that since $\bm{\chi}^\dagger_{\bm{k},n}L^z\bm{\chi}_{\bm{k},n}=0$, only cross terms between $\bm{\chi}_{\bm{k},n}$ can appear in $\langle \mathcal{L}^z\rangle_{\bm{k},T}^{\textrm{shift}} (t)$.
Also, by noting that 
\begin{align}
\bm{\epsilon}_{\bm{k}',L}\cdot \bm{\epsilon}_{\bm{k},T}&=\frac{k_x \delta k_y}{k^2\sqrt{4E_{\bm{k}',L} E_{\bm{k},T}}} \nonumber \\
\bm{\epsilon}_{\bm{k}',T}\cdot \bm{\epsilon}_{\bm{k},T}&=\frac{k^2-k_y \delta k_y}{k^2\sqrt{4 E_{\bm{k}',T} E_{\bm{k},T}}} \label{eq.polarization_dot}
\end{align}
we have
\begin{align}
\alpha^T_{L}&=(E_{\bm{k}',L}+E_{\bm{k},T})\bm{\epsilon}_{\bm{k}',L}\cdot \bm{\epsilon}_{\bm{k},T} \propto \delta k_y \nonumber \\
\alpha^T_{T}&=(E_{\bm{k}',T}+E_{\bm{k},T})\bm{\epsilon}_{\bm{k}',T}\cdot \bm{\epsilon}_{\bm{k},T} \propto (\delta k_y)^0 \nonumber \\
\alpha^T_{-L}&=-(-E_{\bm{k}',L}+E_{\bm{k},T})\bm{\epsilon}_{\bm{k}',L}\cdot \bm{\epsilon}_{\bm{k},T} \propto \delta k_y \nonumber \\ 
\alpha^T_{-T}&=-(-E_{\bm{k}',T}+E_{\bm{k},T})\bm{\epsilon}_{\bm{k}',T}\cdot \bm{\epsilon}_{\bm{k},T} \propto \delta k_y. \label{eq.coefficients}
\end{align}
Thus, to the lowest order in $\delta k_y$, 
\begin{align}
\langle \mathcal{L}^z\rangle_{\bm{k},T}^{\textrm{shift}} (t)=&2\textrm{Re}[\alpha^T_L \alpha^T_T e^{i t (E_{\bm{k}',L}-E_{\bm{k}',T})} \bm{\chi}^\dagger_{\bm{k}',L}L^z \bm{\chi}_{\bm{k}',T}  \nonumber \\ 
&+\alpha^T_T \alpha^T_{-L} e^{it(E_{\bm{k}',T}+E_{\bm{k}',L}) } \bm{\chi}^\dagger_{\bm{k}',T}L^z \bm{\chi}_{\bm{k}',-L}].
\end{align}
Together with
\begin{align}
\bm{\chi}^\dagger_{\bm{k}',L}L^z \bm{\chi}_{\bm{k}',T}&= \frac{i(E_{\bm{k}',L}+E_{\bm{k}',T})}{\sqrt{4E_{\bm{k}',L}E_{\bm{k}',T}} }, \nonumber \\
\bm{\chi}^\dagger_{\bm{k}',T}L^z \bm{\chi}_{\bm{k}',-L}&=\frac{-i(E_{\bm{k}',T}-E_{\bm{k}',L})}{\sqrt{4E_{\bm{k}',L}E_{\bm{k}',T}} } ,
\end{align}
and Eqs.~\eqref{eq.polarization_dot} and \eqref{eq.coefficients}, we see that to the lowest order in $t$ and $\delta k_y$, $\langle \mathcal{L}^z\rangle_{\bm{k},T}^{\textrm{shift}} (t) \propto -k_x \delta k_y t$ with positive coefficients when $v_L>v_T$.
Note that the proportionality to $t$ follows by Taylor expanding the exponential factors.
Similarly, we can show that $\langle \mathcal{L}^z\rangle_{\bm{k},L}^{\textrm{shift}} (t) \propto k_x \delta k_y t$ with positive coefficients when $v_L>v_T$.

We can similarly analyze the three-dimensional continuum with one longitudinal and two transverse modes with energy
\begin{align}
E_{\bm{k},L}&=-E_{\bm{k},-L}=v_L k \\
E_{\bm{k},T_1}&=-E_{\bm{k},-T_1}=v_T k \\
E_{\bm{k},T_2}&=-E_{\bm{k},-T_2}=v_T k 
\end{align}
and polarization
\begin{align}
\bm{\epsilon}_{\bm{k},L}&=\bm{\epsilon}_{\bm{k},-L}=\frac{1}{k\sqrt{2E_{\bm{k},L}}}  \begin{pmatrix}
k_x \\ k_y \\ k_z
\end{pmatrix} \\
\bm{\epsilon}_{\bm{k},T_1}&=\bm{\epsilon}_{\bm{k},-T_1}=\frac{1}{\tilde{k}\sqrt{2E_{\bm{k},T_1}}}\begin{pmatrix}
k_y \\ -k_x \\ 0
\end{pmatrix} \\
\bm{\epsilon}_{\bm{k},T_2}&=\bm{\epsilon}_{\bm{k},-T_2}=\frac{1}{\tilde{k}k\sqrt{2E_{\bm{k},T_2}}}\begin{pmatrix}
k_x k_z \\ k_y k_z \\ -k_x^2-k_y^2
\end{pmatrix},
\end{align}
where we have defined $\tilde{k}=\sqrt{k_x^2+k_y^2}$. 
An analysis parallel to the two-dimensional case shows that the induced angular momentum is given by
\begin{align}
\langle \mathcal{L}^z\rangle_{\bm{k},L}^{\textrm{shift}} (t)&=\frac{k_x \delta k_y t (v_L^2-v_T^2)}{v_L k} \\
\langle \mathcal{L}^z\rangle_{\bm{k},T_1}^{\textrm{shift}} (t)&=\frac{k_x \delta k_y t (v_T^2-v_L^2)}{v_T k} \\
\langle \mathcal{L}^z\rangle_{\bm{k},T_2}^{\textrm{shift}} (t)&=0.
\end{align}

\textbf{Properties of $\bm{\beta^z_{xy}}$}.
Let us begin with the properties of $\beta_{xy}^z$ in the two-dimensional continuum model whose dynamical matrix is given by
\begin{equation}
D_{\bm{k}}=\begin{pmatrix}
v_L^2 k_x^2+v_T^2 k_y^2 & (v_L^2-v_T^2)k_x k_y \\
(v_L^2-v_T^2)k_x k_y & v_T^2 k_x^2+v_L^2 k_y^2
\end{pmatrix}.
\end{equation}
for which the energy and the polarization vectors are given in Eqs.~(\ref{eq.E_L}-\ref{eq.eps_T}).
Omitting the momentum space sector, which vanish, the velocity operators are
\begin{align}
v_x&=\begin{pmatrix}
2 v_L^2 k_x & (v_L^2-v_T^2) k_y \\
(v_L^2-v_T^2) k_y  & 2 v_T^2 k_x
\end{pmatrix} \nonumber \\
v_y&=\begin{pmatrix}
2 v_T^2 k_y & (v_L^2-v_T^2) k_x \\
(v_L^2-v_T^2) k_x  & 2 v_L^2 k_y
\end{pmatrix} .
\end{align}
Recalling the definition of $\Omega_{\mu \nu,n}^{z}(\bm{k} )$, given below Eq.~(3), we find
\begin{align}
\Omega_{xy,L}^{z}(\bm{k} )&= -\frac{2k_x^2(v_L^2+v_T^2)}{k^4(v_L^2-v_T^2)}
\nonumber \\
\Omega_{yx,L}^{z}(\bm{k} )&= \frac{2k_y^2(v_L^2+v_T^2)}{k^4(v_L^2-v_T^2)}
\nonumber \\
\Omega_{xy,T}^{z}(\bm{k} )&=\frac{2k_x^2(v_L^2+v_T^2)}{k^4(v_L^2-v_T^2)}
\nonumber \\
\Omega_{yx,T}^{z}(\bm{k} )&=- \frac{2k_y^2(v_L^2+v_T^2)}{k^4(v_L^2-v_T^2)}
\label{eq.omega}
\end{align}
The point is that $\Omega_{\mu\nu,n}^{z}(\bm{k} )\sim \frac{1}{k^2}$ for $n=L,T$, so that insofar as temperature dependence is concerned, $\beta_{\mu\nu}^z \sim \int dk \frac{1}{k} c_1(g(\frac{v k}{k_BT})) \sim T^{d-2}$.
Thus, we expect $\beta_{\mu\nu}^z$ to approach a constant for $d=2$ and $\beta_{\mu\nu}^z$ to be linear in temperature for $d=3$.
However, for $d=2$, the integral does not converge because $\beta_{\mu\nu}^z \sim \int dk \frac{1}{k} [c_1(g(\frac{v_L k}{k_BT}))-c_1(g(\frac{v_T k}{k_BT})) ]$, and for small $k$, $c_1(g(\frac{v_L k}{k_BT}))-c_1(g(\frac{v_T k}{k_BT}))=O(k^0) $.
Thus, the integral has logarithmic divergence at small $k$.
This divergence is reminiscent of the problem of long longitudinal waves that causes divergence in thermal conductivity \cite{ziman2001electrons}, and indicates that $\beta^z_{xy}$ may depend on the sample size.
PAMHE of two-dimensional systems is studied further in SI \cite{supplement}.

{\small \subsection*{Acknowledgements}
\begin{acknowledgments}
S.P. thanks H.-W. Lee for giving an enlightening talk at MSM19, which motivated this work. S.P. was supported by IBS-R009-D1. B.-J.Y. was supported by the Institute for Basic Science in Korea (Grant No. IBS-R009-D1) and Basic Science Research Program through the National Research Foundation of Korea (NRF) (Grant No. 0426-20190008). This work was supported in part by the U.S. Army Research Office under Grant Number W911NF-18-1-0137.
\end{acknowledgments}}

\end{bibunit}

\begin{bibunit}
\clearpage

\appendix*
\onecolumngrid
\renewcommand{\appendixpagename}{\center \Large Supplementary Information for \\ ``Phonon angular momentum Hall effect''}

\appendixpage
\begin{center} \large Sungjoon Park and Bohm-Jung Yang\end{center}
~\\
\twocolumngrid

\setcounter{equation}{0}
\setcounter{figure}{0}
\setcounter{page}{1}
\renewcommand{\theequation}{{S}\arabic{equation}}
\renewcommand{\thefigure}{{S}\arabic{figure}}

\section{Supplementary Note 1: PAM Hall conductivity}
Here, we review known facts about spin transport in the context of thermal transport of PAM based on Refs.~ \cite{luttinger1964theory,shi2006proper,matsumoto2014thermal,li2019intrinsic}.
Let us write the phonon Hamiltonian as $\mathcal{H}=\frac{1}{2} \int d\bm{r} \bm{x}^\dagger (\bm{r}) \hat{H} \bm{x}(\bm{r})$, where $\hat{H}=\sum_{\bm{\delta}} H_{\bm{\delta}} e^{i \hat{\bm{p}} \cdot \bm{\delta}}$ and $e^{i \hat{\bm{p}} \cdot \bm{\delta}}$ is the translation operator by $\bm{\delta}$.
In the presence of thermal gradient, the phonon Hamiltonian is modified to $\tilde{\mathcal{H}}=\frac{1}{2} \int d\bm{r} \tilde{\bm{x}}^\dagger (\bm{r}) \hat{H} \tilde{\bm{x}}(\bm{r})$, where $\tilde{\bm{x}}(\bm{r})=(1+\frac{\bm{r}\cdot \bm{\nabla} T}{2T})\bm{x}(\bm{r})$.
The PAM current is similarly modified by the temperature gradient, which can be seen by evaluating the time evolution of PAM at position $\bm{r}$ as dictated by the Heisenberg equation $\frac{\partial \bm{\mathcal{L}}(\bm{r})}{dt}=i[\tilde{\mathcal{H}},\bm{\mathcal{L}}(\bm{r})]=-\bm{\nabla}\cdot \bm{j}^{\bm{L}}(\bm{r})+\mathcal{T}^{\bm{L}}(\bm{r})$.
Here,  $\bm{j}^{\bm{L}}(\bm{r})=\tilde{\bm{x}}^{\dagger}(\bm{r}) \frac{\hat{\bm{v}} \sigma^y \bm{L} +\bm{L} \sigma^y \hat{\bm{v}} }{4}\tilde{\bm{x}}(\bm{r}) $ is the local PAM current in the presence of temperature gradient, $\hat{\bm{v}}=i[\hat{H},\bm{r}]$ is the velocity, and $\mathcal{T}^{\bm{L}}(\bm{r})=-\frac{i}{2} \tilde{\bm{x}}^\dagger (\bm{r}) (\bm{L} \sigma^y \hat{H}-\hat{H}\sigma^y \bm{L})\tilde{\bm{x}}(\bm{r})$ is the torque density.
Since the torque density does not vanish in general, continuity equation is not satisfied for this current.
We note in passing that in the presence of inversion symmetry, $\frac{1}{V}\int d\bm{r} \mathcal{T}^{\bm{L}}(\bm{r})=0$ up to the linear order in the temperature gradient.
Introducing the torque dipole density $\mathcal{T}^{\bm{L}}(\bm{r})=-\bm{\nabla}\cdot \bm{p}^{\bm{L}}(\bm{r})$, we have $\mathcal{T}^{\bm{L}}(\bm{k})=-i \bm{k}\cdot \bm{p}^{\bm{L}}(\bm{k})$. 
To the linear order in thermal gradient, we can write $\mathcal{T}^{\bm{L}}(\bm{k})=\bm{\chi}^{\bm{L}}(\bm{k})\cdot \bm{\nabla}T (\bm{k})$, so that the dc response of the torque dipole density is $\bm{p}^{\bm{L}}=\textrm{Re} [i\nabla_{\bm{k}} [\bm{\chi}^{\bm{L}}(\bm{k})\cdot \bm{\nabla}T ] ]_{\bm{k}=0}$.
The current $\mathcal{j}^{\bm{L}}(\bm{r})=\bm{j}^{\bm{L}}(\bm{r})+\bm{p}^{\bm{L}}(\bm{r})$ satisfies the continuity equation and the bulk-boundary correspondence for slowly varying phonon confining potential, i.e. the edge PAM accumulation per area is given by $\mathcal{j}^{\bm{L}}\tau_{\bm{L}}$, where  $\tau_{\bm{L}}$ is the PAM relaxation time.
This bulk-boundary correspondence does not apply for sharp boundaries, and we  focus only on the conventional current.

The PAM Hall conductivity for the conventional current can be obtained by evaluating the expectation value $\langle \bm{j}^{\bm{L}} \rangle_{\textrm{neq}}$ in the presence of temperature gradient, where $\bm{j}^{L^\rho}=\frac{1}{V}\int d\bm{r} \bm{j}^{L^\rho} (\bm{r})=\frac{1}{V}\int d \bm{r} \bm{x}^{\dagger}(\bm{r}) \frac{\hat{\bm{v}} \sigma^y L^\rho +L^\rho \sigma^y \hat{\bm{v}} }{4}\bm{x}(\bm{r}) +\frac{1}{V}\int d \bm{r}  \bm{x}^\dagger (\bm{r})\frac{\{\bm{r}\cdot \bm{\nabla}T ,\hat{\bm{v}} \sigma^y L^\rho +L^\rho \sigma^y \hat{\bm{v}} \}}{8T} \bm{x}(\bm{r})$ and $\{,\}$ is the anticommutator.
In the linear response regime, the non-equilibrium expectation value of the first term can be evaluated using the Kubo formula, while the second term is already proportional to the temperature gradient so that it can be evaluated using the equilibrium distribution.
Evaluating these terms results in Eq.~\eqref{eq.PAMH_conductivity} in the main text.
\section{Supplementary Note 2: Details of the models}
Let us first discuss the simple cubic lattice. 
Define the nearest neighbor spring constant matrix between two atoms separated by $\bm{R}_1$ as
\begin{equation}
K(\bm{R}_1)=\begin{pmatrix}
-k_L & 0 & 0\\
0 & -k_T & 0 \\
0 & 0 & -k_T
\end{pmatrix}. \label{eq.cube_nnspring}
\end{equation}
The other nearest spring constant matrices are given by imposing the symmetry of the simple cubic lattice. 
For example, $K(\bm{R}_2)=C_z(\frac{\pi}{2}) K(\bm{R}_1) C^T_z(\frac{\pi}{2})$, where $C_z(\frac{\pi}{2})$ is the rotation symmetry about the $z$ axis by angle $\frac{\pi}{2}$.
Then, the onsite potential is given by
\begin{equation}
K^{nn}(0)=-2(K(\bm{R}_1)+K(\bm{R}_2)+K(\bm{R}_3)).
\end{equation}
The dynamical matrix for the nearest neighbors is given by
\begin{align}
D^{nn}_{\bm{k}}=&K^{nn}(0)+ 2K(\bm{R}_1)\cos k_x \nonumber \\
&+2K(\bm{R}_2)\cos k_y+2K(\bm{R}_3)\cos k_y.
\end{align}
For nontrivial polarization vectors, we also need the next nearest neighbor spring constants, which be written down in a similar fashion. 
For example, we have
\begin{align}
K(\bm{R}_1+\bm{R}_2)=&C(\tfrac{\pi}{4}) \begin{pmatrix}
-k'_L & 0 & 0\\
0 & -k'_T & 0 \\
0 & 0 & -k'_T
\end{pmatrix} C^T(\tfrac{\pi}{4}) \nonumber \\
=&-\frac{1}{2}
\begin{pmatrix}
k'_{L}+k'_{T} & k'_{L}-k'_{T} & 0\\
k'_{L}-k'_{T} & k'_{L}+k'_{T} & 0 \\
0 & 0 & -k'_{T}
\end{pmatrix},
\end{align}
while the other next nearest neighbor spring constant matrices can be obtained by imposing the symmetry conditions as before.
The onsite potential is similarly given by $K^{nnn}(0)=-\sum_{nnn} K(\Delta \bm{R})$ where $\Delta \bm{R}$ is one of the twelve next nearest sites.
The dynamical matrix for next nearest neighbors is given by
\begin{align}
&D^{nnn}_{\bm{k}}=K^{nnn}(0)+2K(\bm{R}_1+\bm{R}_2) \cos(k_x+k_y) \nonumber \\
&+2K(\bm{R}_1-\bm{R}_2) \cos(k_x-k_y)+2K(\bm{R}_1+\bm{R}_3) \cos(k_x+k_z) \nonumber \\
&+2K(\bm{R}_1-\bm{R}_3) \cos(k_x-k_z) + 2K(\bm{R}_2+\bm{R}_3) \cos(k_y+k_z) \nonumber \\
&+2K(\bm{R}_2-\bm{R}_3) \cos(k_x-k_z).
\end{align}

In the case of CsCl lattice, we have two sites, $A$ and $B$, per unit cell.   
Define $\bm{\delta}_A=0$, $\bm{\delta}_B=\frac{\bm{R}_1+\bm{R}_2+\bm{R}_3}{2}$.
The nearest spring constant between A and B sites separated by $\bm{\delta}_{B}$ is given by
\begin{equation}
K(\bm{\delta}_B)=-\frac{1}{3}\begin{pmatrix}
k_L+2k_T& k_L-k_T&  k_L-k_T \\ 
k_L-k_T & k_L+2k_T & k_L-k_T \\
k_L-k_T & k_L-k_T & k_L+2k_T
\end{pmatrix},
\end{equation}
while the others can be obtained by imposing the lattice symmetries.
The next nearest neighbor spring constant matrix separated by $\bm{R}_1$
\begin{equation}
K^\alpha(\bm{R}_1)=\begin{pmatrix}
-k^\alpha_L & 0 & 0\\
0 & -k^\alpha_T & 0 \\
0 & 0 & -k^\alpha_T
\end{pmatrix},
\end{equation}
where $\alpha=A$ or $B$, while the others are obtained by imposing the lattice symmetries. 
As before, the onsite potential is obtained by demanding that there is no energy cost in uniform translation of the lattice:
$K^{\alpha}_{nn}(0)=-2(K(\bm{\delta}_B)+K(\bm{\delta}_B-\bm{R}_1)+K(\bm{\delta}_B-\bm{R}_2)+K(\bm{\delta}_B-\bm{R}_3))$, and $K^{\alpha}_{nnn}(0)=-2K^{\alpha}(\bm{R}_1)-2K^{\alpha}(\bm{R}_2)-2K^{\alpha}(\bm{R}_3)$.
The dynamical matrix takes the form
\begin{equation}
D_{\bm{k}}=\begin{pmatrix}
D^A_{\bm{k}} & D^{AB}_{\bm{k}}\\
D^{AB}_{\bm{k}} & D^B_{\bm{k}}
\end{pmatrix},
\end{equation}
where $D^\alpha_{\bm{k}}=K^{\alpha}_{nn}(0)+K^{\alpha}_{nnn}(0)+2K^{\alpha}(\bm{R}_1)\cos k_x+2K^{\alpha}(\bm{R}_2)\cos k_y+2K^{\alpha}(\bm{R}_3)\cos k_z$ and $D^{AB}_{\bm{k}}=2K(\bm{\delta}_B) \cos \frac{k_x+k_y+k_z}{2}+2K(\bm{\delta}_B-\bm{R}_1) \cos \frac{-k_x+k_y+k_z}{2}+2K(\bm{\delta}_B-\bm{R}_2) \cos \frac{k_x-k_y+k_z}{2}+2K(\bm{\delta}_B-\bm{R}_3) \cos \frac{k_x+k_y-k_z}{2}$.
It is easy to see that the process of interchanging $k^A_{L},k^A_{T} \leftrightarrow k^B_{L},k^B_{T}$ is equivalent to the transformation $D_{\bm{k}}\rightarrow D'_{\bm{k}}=I_{AB} D_{\bm{k}} I_{AB}^\dagger$, where 
\begin{equation}
I_{AB}=\begin{pmatrix}
0 & I_3 \\
I_3 & 0
\end{pmatrix}
\end{equation}
and $I_3$ is the three by three identity matrix.
Thus, if $D_{\bm{k}}\bm{\epsilon}_{\bm{k},n}=E^2_{\bm{k},n}\bm{\epsilon}_{\bm{k},n}$, $D'_{\bm{k}}\bm{\epsilon}'_{\bm{k},n}=E^2_{\bm{k},n}\bm{\epsilon}'_{\bm{k},n}$, where $\bm{\epsilon}'_{\bm{k},n}=I_{AB}\bm{\epsilon}_{\bm{k},n}$.
Thus, under the interchange of the spring constant values, the energy spectrum is not changed in any way.
Noting that the phonon wavefunction is given by $\bm{\chi}_{\bm{k},n}=(\begin{smallmatrix}-iE_{\bm{k},n}\bm{\epsilon}_{\bm{k},n} \\ \bm{\epsilon}_{\bm{k},n} \end{smallmatrix})$, we see that the wavefunction content of $A$ and $B$ are interchanged.

\section{Supplementary Note 3: Two-dimensional lattice models}

\begin{figure*}[t]
\centering
\includegraphics[width=17cm]{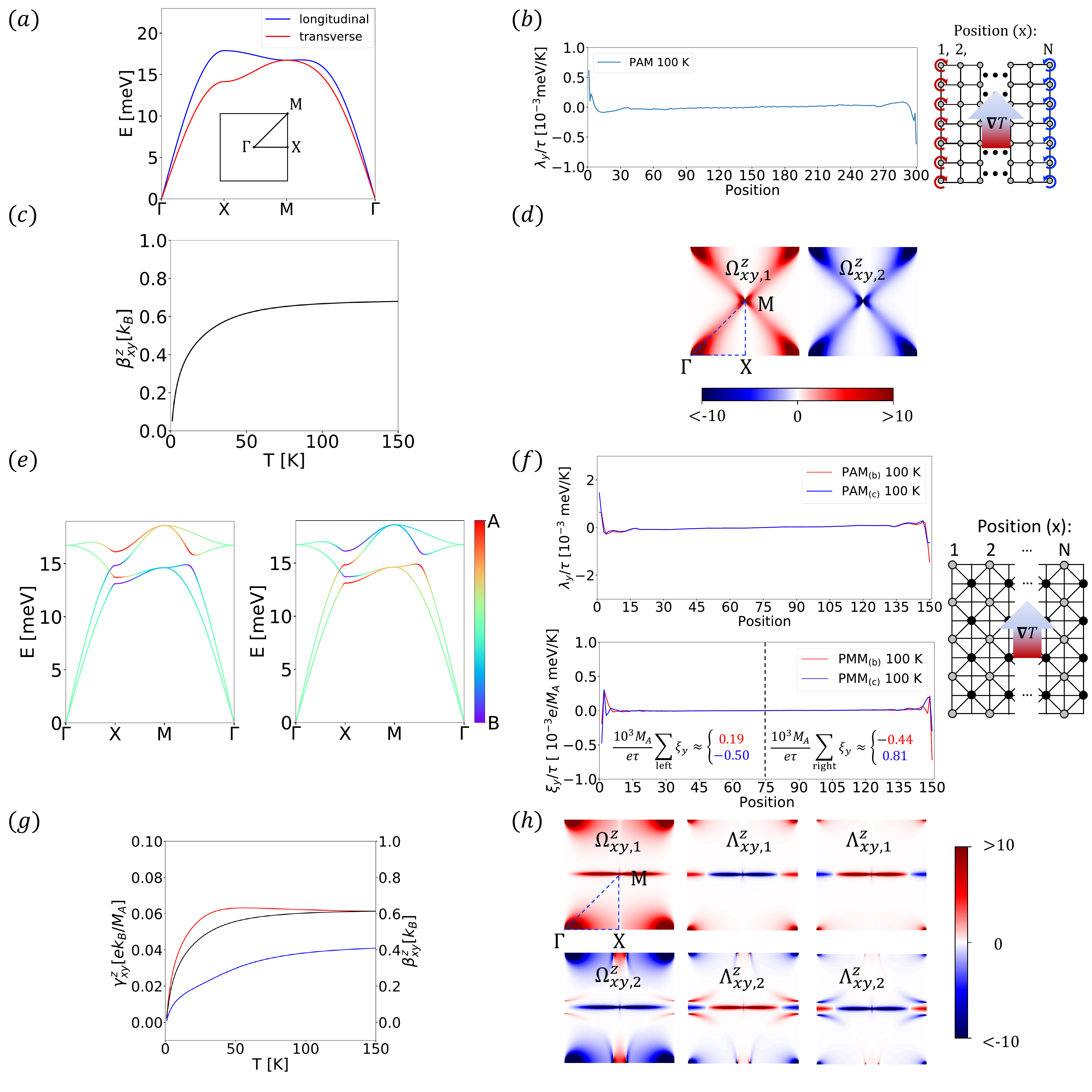}
\caption{\textbf{Study of two-dimensional lattices.} 
(a) The phonon energy spectrum of the square lattice for the parameters in Fig.~\ref{fig.cube} (a). 
(b) PAM distribution for the square lattice with edges. 
(c) PAM Hall conductivity for square lattice.
(d) $\Omega^z_{xy,1}$ (left) and $\Omega^z_{xy,2}$ (right) for the square lattice.
(e) Left and right figures show the energy spectrum of the checkerboard lattice with the parameters in Fig.~\ref{fig.cscl} (b) and (c), respectively. The color represents the wavefunction content.
(f) Upper figure shows the PAM distribution for the checkerboard lattice with edges using the parameters in Fig.~\ref{fig.cscl} (b) (red curve) and Fig.~\ref{fig.cscl} (c) (blue curve), and the lower figure shows PMM distribution for the checkerboard lattice with edges using the parameters in Fig.~\ref{fig.cscl} (b) (red curve) and Fig.~\ref{fig.cscl} (c) (blue curve).
(g) PAM and PMM Hall conductivities for checkerboard lattice. 
The black line is $\beta^z_{xy}$ and the red and the blue lines are $\gamma^z_{xy}$ for parameters in  Fig.~\ref{fig.cscl} (b) (red) and Fig.~\ref{fig.cscl} (c) (blue) respectively.
(h) Leftmost pair shows $\Omega^z_{xy,1}$ (upper) and $\Omega^z_{xy,2}$ (lower) for the checkerboard lattice, the middle pair shows $\Lambda^z_{xy,1}$ (upper) and $\Lambda^z_{xy,2}$ (lower) computed using the parameters in Fig.~\ref{fig.cscl} (b), and the rightmost pair shows  $\Lambda^z_{xy,1}$ (upper) and $\Lambda^z_{xy,2}$ (lower) computed using the parameters in Fig.~\ref{fig.cscl} (c).
}
\label{fig.2d_models}
\end{figure*}

In this section, we examine the two-dimensional square lattice and checkerboard lattice models.
Let us begin with the square lattice. 
We have included only the nearest and next nearest spring constants with the values given in Fig.~\ref{fig.cube} in the main text, and the energy spectrum of the model is shown in Fig.~\ref{fig.2d_models} (a).
Although the PAM Hall conductivity $\beta^z_{xy}$ calculated using Eq.~\eqref{eq.PAMH_conductivity} in the main text diverges for infinitely large system, there is no divergence in the PAM accumulation $\lambda_{y}(x)$ induced by thermal gradient, which we show in Fig.~\ref{fig.2d_models} (b).
This suggests that the PAM Hall current does not diverge in a finite size system.
Therefore, we approximate the finite size effect by introducing a cutoff  in the summation over $\bm{k}$  when calculating $\beta^z_{xy}$ by restricting $|k_x| > \frac{2\pi}{L}$, where $L=300$ is the length along the $x$ direction, which we show in Fig.~\ref{fig.2d_models} (c).
For reference, we show the PAM curvature  $\Omega_{xy,n}^z(\bm{k})$ in Fig.~\ref{fig.2d_models} (d).

Let us note that the dependence of $\beta^z_{xy}$ on the size of the system suggests that PAM accumulation may be sensitive to the details of the boundary.
In addition, because PAM is not conserved, the PAM torque dipole density is also expected to play an important role in the PAMHE, whose evaluation we leave for future work.

Next, we discuss the checkerboard lattice.
Following the discussion for the CsCl lattice model in the main text, we use the two sets of parameters in Fig.~\ref{fig.cscl} to plot the energy and the wavefunction content in Fig.~\ref{fig.2d_models} (e).
Specifically, the parameters used to plot Fig.~\ref{fig.cscl} (b) is used to plot Fig.~\ref{fig.2d_models} (e)  (left) and the parameters used to plot Fig.~\ref{fig.cscl} (c) is used to plot Fig.~\ref{fig.2d_models} (e) (right). 
Similar to the case for the CsCl lattice, the energy spectrum is the same for the two sets of parameters while the wave function contents change.

We show the PAM and PMM accumulations ($\lambda_{y}(x)$ and $\xi_{y}(x)$, respectively) in Fig.~\ref{fig.2d_models} (f) (upper and lower, respectively).
There are no divergences in the PAM and PMM accumulations, although the PAM and PMM Hall conductivities ($\beta^z_{xy}$ and $\gamma^z_{xy}$ respectively) diverge for infinite size systems.
We thus introduce a cutoff $|k_x| > \frac{2\pi}{L}$ with $L=150$ to calculate them, which are shown in Fig.~\ref{fig.2d_models} (g).
We note that the behaviors of the PMM conductivities can be analyzed as in main text for CsCl lattice.
We can also expect that the PAM and PMM accumulations are sensitive to the details of the boundary, and that torque dipole density can be important for PAMHE and PMMHE.

To help understand the behavior of the conductivities, it is useful to plot the PAM and PMM curvatures ($\Omega_{xy,n}^z(\bm{k})$ and $\Lambda_{xy,n}^z(\bm{k})$ respectively), which are shown in Fig.~\ref{fig.2d_models} (h).
As the low energy modes are important in the temperature region we consider, only the curvatures of the lowest two bands are shown.
The leftmost pair shows $\Omega^z_{xy, 1}(\bm{k})$ (upper) and $\Omega^z_{xy, 2}(\bm{k})$ (lower) for the checkerboard lattice (note that the energy bands are labeled from 4 (highest energy) to 1 (lowest energy). 
Let us note that as in the CsCl lattice model, the PAM curvature for the two sets of parameters are equivalent.
The middle pair of  $\Lambda^z_{xy,n}(\bm{k})$ in Fig.~\ref{fig.2d_models} (h) is computed using the parameters in Fig.~\ref{fig.cscl} (b), and the rightmost pair of $\Lambda^z_{xy,n}(\bm{k})$ in Fig.~\ref{fig.2d_models} (h) is computed using the parameters in Fig.~\ref{fig.cscl} (c).

Finally, let us note that the induced PMM near the edges has order of magnitude of $\tau/(1s)$ Bohr magneton per unit cell for the blue curve, where we have assumed that the unit cell size is $5\textrm{\AA}$, $M_A$ is $25$ amu, and the temperature gradient is $10$K/$100\mu$m.
If we assume that the acoustic phonon lifetime is $10\sim 100$ ps, the PMM accumulation can be expected to be around $10^{-11}\sim10^{-10}$ Bohr magneton per unit cell.
This value is similar to the bulk PMM induced by the thermal analog of Edelstein effect obtained in Ref.~\cite{hamada2018phonon} for $\textrm{GaN}$.

\end{bibunit}

\begin{thebibliography}{41}
\expandafter\ifx\csname natexlab\endcsname\relax\def\natexlab#1{#1}\fi
\expandafter\ifx\csname bibnamefont\endcsname\relax
  \def\bibnamefont#1{#1}\fi
\expandafter\ifx\csname bibfnamefont\endcsname\relax
  \def\bibfnamefont#1{#1}\fi
\expandafter\ifx\csname citenamefont\endcsname\relax
  \def\citenamefont#1{#1}\fi
\expandafter\ifx\csname url\endcsname\relax
  \def\url#1{\texttt{#1}}\fi
\expandafter\ifx\csname urlprefix\endcsname\relax\def\urlprefix{URL }\fi
\providecommand{\bibinfo}[2]{#2}
\providecommand{\eprint}[2][]{\url{#2}}

\bibitem[{\citenamefont{Nagaosa et~al.}(2010)\citenamefont{Nagaosa, Sinova,
  Onoda, MacDonald, and Ong}}]{nagaosa2010anomalous}
\bibinfo{author}{\bibfnamefont{N.}~\bibnamefont{Nagaosa}},
  \bibinfo{author}{\bibfnamefont{J.}~\bibnamefont{Sinova}},
  \bibinfo{author}{\bibfnamefont{S.}~\bibnamefont{Onoda}},
  \bibinfo{author}{\bibfnamefont{A.~H.} \bibnamefont{MacDonald}},
  \bibnamefont{and} \bibinfo{author}{\bibfnamefont{N.~P.} \bibnamefont{Ong}},
  \bibinfo{journal}{Rev. Mod. Phys.} \textbf{\bibinfo{volume}{82}},
  \bibinfo{pages}{1539} (\bibinfo{year}{2010}).
  
  \bibitem[{\citenamefont{Sinova et~al.}(2015)\citenamefont{Sinova, Valenzuela,
  Wunderlich, Back, and Jungwirth}}]{sinova2015spin}
\bibinfo{author}{\bibfnamefont{J.}~\bibnamefont{Sinova}},
  \bibinfo{author}{\bibfnamefont{S.~O.} \bibnamefont{Valenzuela}},
  \bibinfo{author}{\bibfnamefont{J.}~\bibnamefont{Wunderlich}},
  \bibinfo{author}{\bibfnamefont{C.~H.}~\bibnamefont{Back}}, \bibnamefont{and}
  \bibinfo{author}{\bibfnamefont{T.}~\bibnamefont{Jungwirth}},
  \bibinfo{journal}{Rev. Mod. Phys.} \textbf{\bibinfo{volume}{87}},
  \bibinfo{pages}{1213} (\bibinfo{year}{2015}).
  
\bibitem[{\citenamefont{Karplus and Luttinger}(1954)}]{karplus1954hall}
\bibinfo{author}{\bibfnamefont{R.}~\bibnamefont{Karplus}} \bibnamefont{and}
  \bibinfo{author}{\bibfnamefont{J.}~\bibnamefont{Luttinger}},
  \bibinfo{journal}{Phys. Rev.} \textbf{\bibinfo{volume}{95}},
  \bibinfo{pages}{1154} (\bibinfo{year}{1954}).

\bibitem[{\citenamefont{Smit}(1958)}]{smit1958spontaneous}
\bibinfo{author}{\bibfnamefont{J.}~\bibnamefont{Smit}},
  \bibinfo{journal}{Physica} \textbf{\bibinfo{volume}{24}}, \bibinfo{pages}{39}
  (\bibinfo{year}{1958}).

\bibitem[{\citenamefont{Berger}(1970)}]{berger1970side}
\bibinfo{author}{\bibfnamefont{L.}~\bibnamefont{Berger}},
  \bibinfo{journal}{Phys. Rev. B} \textbf{\bibinfo{volume}{2}},
  \bibinfo{pages}{4559} (\bibinfo{year}{1970}).

\bibitem[{\citenamefont{D'yakonov and Perel}(1971)}]{d1971possibility}
\bibinfo{author}{\bibfnamefont{M.}~\bibnamefont{D'yakonov}} \bibnamefont{and}
  \bibinfo{author}{\bibfnamefont{V.}~\bibnamefont{Perel}},
  \bibinfo{journal}{JETP Lett.} \textbf{\bibinfo{volume}{13}},
  \bibinfo{pages}{467} (\bibinfo{year}{1971}).

\bibitem[{\citenamefont{Hirsch}(1999)}]{hirsch1999spin}
\bibinfo{author}{\bibfnamefont{J.~E.}~\bibnamefont{Hirsch}},
  \bibinfo{journal}{Phys. Rev. Lett.} \textbf{\bibinfo{volume}{83}},
  \bibinfo{pages}{1834} (\bibinfo{year}{1999}).

\bibitem[{\citenamefont{Murakami et~al.}(2003)\citenamefont{Murakami, Nagaosa,
  and Zhang}}]{murakami2003dissipationless}
\bibinfo{author}{\bibfnamefont{S.}~\bibnamefont{Murakami}},
  \bibinfo{author}{\bibfnamefont{N.}~\bibnamefont{Nagaosa}}, \bibnamefont{and}
  \bibinfo{author}{\bibfnamefont{S.-C.} \bibnamefont{Zhang}},
  \bibinfo{journal}{Science} \textbf{\bibinfo{volume}{301}},
  \bibinfo{pages}{1348} (\bibinfo{year}{2003}).

\bibitem[{\citenamefont{Sinova et~al.}(2004)\citenamefont{Sinova, Culcer, Niu,
  Sinitsyn, Jungwirth, and MacDonald}}]{sinova2004universal}
\bibinfo{author}{\bibfnamefont{J.}~\bibnamefont{Sinova}},
  \bibinfo{author}{\bibfnamefont{D.}~\bibnamefont{Culcer}},
  \bibinfo{author}{\bibfnamefont{Q.}~\bibnamefont{Niu}},
  \bibinfo{author}{\bibfnamefont{N.~A.}~\bibnamefont{Sinitsyn}},
  \bibinfo{author}{\bibfnamefont{T.}~\bibnamefont{Jungwirth}},
  \bibnamefont{and} \bibinfo{author}{\bibfnamefont{A.~H.}
  \bibnamefont{MacDonald}}, \bibinfo{journal}{Phys. Rev. Lett.}
  \textbf{\bibinfo{volume}{92}}, \bibinfo{pages}{126603}
  (\bibinfo{year}{2004}).

\bibitem[{\citenamefont{Kato et~al.}(2004)\citenamefont{Kato, Myers, Gossard,
  and Awschalom}}]{kato2004observation}
\bibinfo{author}{\bibfnamefont{Y.~K.} \bibnamefont{Kato}},
  \bibinfo{author}{\bibfnamefont{R.~C.} \bibnamefont{Myers}},
  \bibinfo{author}{\bibfnamefont{A.~C.} \bibnamefont{Gossard}},
  \bibnamefont{and} \bibinfo{author}{\bibfnamefont{D.~D.}
  \bibnamefont{Awschalom}}, \bibinfo{journal}{Science}
  \textbf{\bibinfo{volume}{306}}, \bibinfo{pages}{1910} (\bibinfo{year}{2004}).

\bibitem[{\citenamefont{Wunderlich et~al.}(2005)\citenamefont{Wunderlich,
  Kaestner, Sinova, and Jungwirth}}]{wunderlich2005experimental}
\bibinfo{author}{\bibfnamefont{J.}~\bibnamefont{Wunderlich}},
  \bibinfo{author}{\bibfnamefont{B.}~\bibnamefont{Kaestner}},
  \bibinfo{author}{\bibfnamefont{J.}~\bibnamefont{Sinova}}, \bibnamefont{and}
  \bibinfo{author}{\bibfnamefont{T.}~\bibnamefont{Jungwirth}},
  \bibinfo{journal}{Phys. Rev. Lett.} \textbf{\bibinfo{volume}{94}},
  \bibinfo{pages}{047204} (\bibinfo{year}{2005}).

\bibitem[{\citenamefont{Bernevig et~al.}(2005)\citenamefont{Bernevig, Hughes,
  and Zhang}}]{bernevig2005orbitronics}
\bibinfo{author}{\bibfnamefont{B.~A.} \bibnamefont{Bernevig}},
  \bibinfo{author}{\bibfnamefont{T.~L.} \bibnamefont{Hughes}},
  \bibnamefont{and} \bibinfo{author}{\bibfnamefont{S.-C.} \bibnamefont{Zhang}},
  \bibinfo{journal}{Phys. Rev. Lett.} \textbf{\bibinfo{volume}{95}},
  \bibinfo{pages}{066601} (\bibinfo{year}{2005}).

\bibitem[{\citenamefont{Haldane}(1988)}]{haldane1988model}
\bibinfo{author}{\bibfnamefont{F.~D.~M.} \bibnamefont{Haldane}},
  \bibinfo{journal}{Phys. Rev. Lett.} \textbf{\bibinfo{volume}{61}},
  \bibinfo{pages}{2015} (\bibinfo{year}{1988}).

\bibitem[{\citenamefont{Kane and Mele}(2005{\natexlab{a}})}]{kane2005quantum}
\bibinfo{author}{\bibfnamefont{C.~L.} \bibnamefont{Kane}} \bibnamefont{and}
  \bibinfo{author}{\bibfnamefont{E.~J.} \bibnamefont{Mele}},
  \bibinfo{journal}{Phys. Rev. Lett.} \textbf{\bibinfo{volume}{95}},
  \bibinfo{pages}{226801} (\bibinfo{year}{2005}{\natexlab{a}}).

\bibitem[{\citenamefont{Kane and Mele}(2005{\natexlab{b}})}]{kane2005z}
\bibinfo{author}{\bibfnamefont{C.~L.} \bibnamefont{Kane}} \bibnamefont{and}
  \bibinfo{author}{\bibfnamefont{E.~J.} \bibnamefont{Mele}},
  \bibinfo{journal}{Phys. Rev. Lett.} \textbf{\bibinfo{volume}{95}},
  \bibinfo{pages}{146802} (\bibinfo{year}{2005}{\natexlab{b}}).

\bibitem[{\citenamefont{Popovi{\'c}}(1989)}]{popovic1989hall}
\bibinfo{author}{\bibfnamefont{R.}~\bibnamefont{Popovi{\'c}}},
  \bibinfo{journal}{Sensors and Actuators} \textbf{\bibinfo{volume}{17}},
  \bibinfo{pages}{39} (\bibinfo{year}{1989}).

\bibitem[{\citenamefont{Jungwirth et~al.}(2012)\citenamefont{Jungwirth,
  Wunderlich, and Olejn{\'\i}k}}]{jungwirth2012spin}
\bibinfo{author}{\bibfnamefont{T.}~\bibnamefont{Jungwirth}},
  \bibinfo{author}{\bibfnamefont{J.}~\bibnamefont{Wunderlich}},
  \bibnamefont{and}
  \bibinfo{author}{\bibfnamefont{K.}~\bibnamefont{Olejn{\'\i}k}},
  \bibinfo{journal}{Nat. Mater.} \textbf{\bibinfo{volume}{11}},
  \bibinfo{pages}{382} (\bibinfo{year}{2012}).
  
\bibitem[{\citenamefont{Katsura et~al.}(2010)\citenamefont{Katsura, Nagaosa,
  and Lee}}]{katsura2010theory}
\bibinfo{author}{\bibfnamefont{H.}~\bibnamefont{Katsura}},
  \bibinfo{author}{\bibfnamefont{N.}~\bibnamefont{Nagaosa}}, \bibnamefont{and}
  \bibinfo{author}{\bibfnamefont{P.~A.} \bibnamefont{Lee}},
  \bibinfo{journal}{Phys. Rev. Lett.} \textbf{\bibinfo{volume}{104}},
  \bibinfo{pages}{066403} (\bibinfo{year}{2010}).

\bibitem[{\citenamefont{Onose et~al.}(2010)\citenamefont{Onose, Ideue, Katsura,
  Shiomi, Nagaosa, and Tokura}}]{onose2010observation}
\bibinfo{author}{\bibfnamefont{Y.}~\bibnamefont{Onose}},
  \bibinfo{author}{\bibfnamefont{T.}~\bibnamefont{Ideue}},
  \bibinfo{author}{\bibfnamefont{H.}~\bibnamefont{Katsura}},
  \bibinfo{author}{\bibfnamefont{Y.}~\bibnamefont{Shiomi}},
  \bibinfo{author}{\bibfnamefont{N.}~\bibnamefont{Nagaosa}}, \bibnamefont{and}
  \bibinfo{author}{\bibfnamefont{Y.}~\bibnamefont{Tokura}},
  \bibinfo{journal}{Science} \textbf{\bibinfo{volume}{329}},
  \bibinfo{pages}{297} (\bibinfo{year}{2010}).

\bibitem[{\citenamefont{Matsumoto and
  Murakami}(2011)}]{matsumoto2011theoretical}
\bibinfo{author}{\bibfnamefont{R.}~\bibnamefont{Matsumoto}} \bibnamefont{and}
  \bibinfo{author}{\bibfnamefont{S.}~\bibnamefont{Murakami}},
  \bibinfo{journal}{Phys. Rev. Lett.} \textbf{\bibinfo{volume}{106}},
  \bibinfo{pages}{197202} (\bibinfo{year}{2011}).

\bibitem[{\citenamefont{Strohm et~al.}(2005)\citenamefont{Strohm, Rikken, and
  Wyder}}]{strohm2005phenomenological}
\bibinfo{author}{\bibfnamefont{C.}~\bibnamefont{Strohm}},
  \bibinfo{author}{\bibfnamefont{G.~L.~J.~A.}~\bibnamefont{Rikken}}, \bibnamefont{and}
  \bibinfo{author}{\bibfnamefont{P.}~\bibnamefont{Wyder}},
  \bibinfo{journal}{Phys. Rev. Lett.} \textbf{\bibinfo{volume}{95}},
  \bibinfo{pages}{155901} (\bibinfo{year}{2005}).

\bibitem[{\citenamefont{Sheng et~al.}(2006)\citenamefont{Sheng, Sheng, and
  Ting}}]{sheng2006theory}
\bibinfo{author}{\bibfnamefont{L.}~\bibnamefont{Sheng}},
  \bibinfo{author}{\bibfnamefont{D.~N.}~\bibnamefont{Sheng}}, \bibnamefont{and}
  \bibinfo{author}{\bibfnamefont{C.~S.}~\bibnamefont{Ting}},
  \bibinfo{journal}{Phys. Rev. Lett.} \textbf{\bibinfo{volume}{96}},
  \bibinfo{pages}{155901} (\bibinfo{year}{2006}).

\bibitem[{\citenamefont{Kagan and Maksimov}(2008)}]{kagan2008anomalous}
\bibinfo{author}{\bibfnamefont{Y.}~\bibnamefont{Kagan}} \bibnamefont{and}
  \bibinfo{author}{\bibfnamefont{L.~A.}~\bibnamefont{Maksimov}},
  \bibinfo{journal}{Phys. Rev. Lett.} \textbf{\bibinfo{volume}{100}},
  \bibinfo{pages}{145902} (\bibinfo{year}{2008}).

\bibitem[{\citenamefont{Zhang et~al.}(2010)\citenamefont{Zhang, Ren, Wang, and
  Li}}]{zhang2010topological}
\bibinfo{author}{\bibfnamefont{L.}~\bibnamefont{Zhang}},
  \bibinfo{author}{\bibfnamefont{J.}~\bibnamefont{Ren}},
  \bibinfo{author}{\bibfnamefont{J.-S.} \bibnamefont{Wang}}, \bibnamefont{and}
  \bibinfo{author}{\bibfnamefont{B.}~\bibnamefont{Li}}, \bibinfo{journal}{Phys.
  Rev. Lett.} \textbf{\bibinfo{volume}{105}}, \bibinfo{pages}{225901}
  (\bibinfo{year}{2010}).

\bibitem[{\citenamefont{Park and Yang}(2019)}]{park2019topological}
\bibinfo{author}{\bibfnamefont{S.}~\bibnamefont{Park}} \bibnamefont{and}
  \bibinfo{author}{\bibfnamefont{B.-J.} \bibnamefont{Yang}},
  \bibinfo{journal}{Phys. Rev. B} \textbf{\bibinfo{volume}{99}},
  \bibinfo{pages}{174435} (\bibinfo{year}{2019}).

\bibitem[{\citenamefont{Zhang et~al.}(2019{\natexlab{a}})\citenamefont{Zhang,
  Zhang, Okamoto, and Xiao}}]{zhang2019thermal}
\bibinfo{author}{\bibfnamefont{X.}~\bibnamefont{Zhang}},
  \bibinfo{author}{\bibfnamefont{Y.}~\bibnamefont{Zhang}},
  \bibinfo{author}{\bibfnamefont{S.}~\bibnamefont{Okamoto}}, \bibnamefont{and}
  \bibinfo{author}{\bibfnamefont{D.}~\bibnamefont{Xiao}},
  \bibinfo{journal}{Phys. Rev. Lett.} \textbf{\bibinfo{volume}{123}},
  \bibinfo{pages}{167202} (\bibinfo{year}{2019}{\natexlab{a}}).

\bibitem[{\citenamefont{Cheng et~al.}(2016)\citenamefont{Cheng, Okamoto, and
  Xiao}}]{cheng2016spin}
\bibinfo{author}{\bibfnamefont{R.}~\bibnamefont{Cheng}},
  \bibinfo{author}{\bibfnamefont{S.}~\bibnamefont{Okamoto}}, \bibnamefont{and}
  \bibinfo{author}{\bibfnamefont{D.}~\bibnamefont{Xiao}},
  \bibinfo{journal}{Phys. Rev. Lett.} \textbf{\bibinfo{volume}{117}},
  \bibinfo{pages}{217202} (\bibinfo{year}{2016}).

\bibitem[{\citenamefont{Zyuzin and Kovalev}(2016)}]{zyuzin2016magnon}
\bibinfo{author}{\bibfnamefont{V.~A.} \bibnamefont{Zyuzin}} \bibnamefont{and}
  \bibinfo{author}{\bibfnamefont{A.~A.} \bibnamefont{Kovalev}},
  \bibinfo{journal}{Phys. Rev. Lett.} \textbf{\bibinfo{volume}{117}},
  \bibinfo{pages}{217203} (\bibinfo{year}{2016}).

\bibitem[{\citenamefont{Park et~al.}(2019)\citenamefont{Park, Nagaosa, and
  Yang}}]{park2019thermal}
\bibinfo{author}{\bibfnamefont{S.}~\bibnamefont{Park}},
  \bibinfo{author}{\bibfnamefont{N.}~\bibnamefont{Nagaosa}}, \bibnamefont{and}
  \bibinfo{author}{\bibfnamefont{B.-J.} \bibnamefont{Yang}},
  \bibinfo{journal}{arXiv preprint arXiv:1910.07206}  (\bibinfo{year}{2019}).

\bibitem[{\citenamefont{Zhang et~al.}(2019{\natexlab{b}})\citenamefont{Zhang,
  Go, Lee, and Kim}}]{zhang20193}
\bibinfo{author}{\bibfnamefont{S.}~\bibnamefont{Zhang}},
  \bibinfo{author}{\bibfnamefont{G.}~\bibnamefont{Go}},
  \bibinfo{author}{\bibfnamefont{K.-J.} \bibnamefont{Lee}}, \bibnamefont{and}
  \bibinfo{author}{\bibfnamefont{S.~K.} \bibnamefont{Kim}},
  \bibinfo{journal}{arXiv preprint arXiv:1909.08031}
  (\bibinfo{year}{2019}{\natexlab{b}}).

\bibitem[{\citenamefont{Zhang and Niu}(2014)}]{zhang2014angular}
\bibinfo{author}{\bibfnamefont{L.}~\bibnamefont{Zhang}} \bibnamefont{and}
  \bibinfo{author}{\bibfnamefont{Q.}~\bibnamefont{Niu}},
  \bibinfo{journal}{Phys. Rev. Lett.} \textbf{\bibinfo{volume}{112}},
  \bibinfo{pages}{085503} (\bibinfo{year}{2014}).

\bibitem[{\citenamefont{Juraschek et~al.}(2017)\citenamefont{Juraschek,
  Fechner, Balatsky, and Spaldin}}]{juraschek2017dynamical}
\bibinfo{author}{\bibfnamefont{D.~M.} \bibnamefont{Juraschek}},
  \bibinfo{author}{\bibfnamefont{M.}~\bibnamefont{Fechner}},
  \bibinfo{author}{\bibfnamefont{A.~V.} \bibnamefont{Balatsky}},
  \bibnamefont{and} \bibinfo{author}{\bibfnamefont{N.~A.}
  \bibnamefont{Spaldin}}, \bibinfo{journal}{Phys. Rev. Mater.}
  \textbf{\bibinfo{volume}{1}}, \bibinfo{pages}{014401} (\bibinfo{year}{2017}).

\bibitem[{\citenamefont{Hamada et~al.}(2018)\citenamefont{Hamada, Minamitani,
  Hirayama, and Murakami}}]{hamada2018phonon}
\bibinfo{author}{\bibfnamefont{M.}~\bibnamefont{Hamada}},
  \bibinfo{author}{\bibfnamefont{E.}~\bibnamefont{Minamitani}},
  \bibinfo{author}{\bibfnamefont{M.}~\bibnamefont{Hirayama}}, \bibnamefont{and}
  \bibinfo{author}{\bibfnamefont{S.}~\bibnamefont{Murakami}},
  \bibinfo{journal}{Phys. Rev. Lett.} \textbf{\bibinfo{volume}{121}},
  \bibinfo{pages}{175301} (\bibinfo{year}{2018}).

\bibitem[{\citenamefont{Juraschek and Spaldin}(2019)}]{juraschek2019orbital}
\bibinfo{author}{\bibfnamefont{D.~M.} \bibnamefont{Juraschek}}
  \bibnamefont{and} \bibinfo{author}{\bibfnamefont{N.~A.}
  \bibnamefont{Spaldin}}, \bibinfo{journal}{Phys. Rev. Mater.}
  \textbf{\bibinfo{volume}{3}}, \bibinfo{pages}{064405} (\bibinfo{year}{2019}).

\bibitem[{\citenamefont{Go et~al.}(2018)\citenamefont{Go, Jo, Kim, and
  Lee}}]{go2018intrinsic}
\bibinfo{author}{\bibfnamefont{D.}~\bibnamefont{Go}},
  \bibinfo{author}{\bibfnamefont{D.}~\bibnamefont{Jo}},
  \bibinfo{author}{\bibfnamefont{C.}~\bibnamefont{Kim}}, \bibnamefont{and}
  \bibinfo{author}{\bibfnamefont{H.-W.} \bibnamefont{Lee}},
  \bibinfo{journal}{Phys. Rev. Lett.} \textbf{\bibinfo{volume}{121}},
  \bibinfo{pages}{086602} (\bibinfo{year}{2018}).

\bibitem[{\citenamefont{Luttinger}(1964)}]{luttinger1964theory}
\bibinfo{author}{\bibfnamefont{J.}~\bibnamefont{Luttinger}},
  \bibinfo{journal}{Phys. Rev.} \textbf{\bibinfo{volume}{135}},
  \bibinfo{pages}{A1505} (\bibinfo{year}{1964}).

\bibitem[{\citenamefont{Matsumoto et~al.}(2014)\citenamefont{Matsumoto,
  Shindou, and Murakami}}]{matsumoto2014thermal}
\bibinfo{author}{\bibfnamefont{R.}~\bibnamefont{Matsumoto}},
  \bibinfo{author}{\bibfnamefont{R.}~\bibnamefont{Shindou}}, \bibnamefont{and}
  \bibinfo{author}{\bibfnamefont{S.}~\bibnamefont{Murakami}},
  \bibinfo{journal}{Phys. Rev. B} \textbf{\bibinfo{volume}{89}},
  \bibinfo{pages}{054420} (\bibinfo{year}{2014}).

\bibitem[{\citenamefont{Li et~al.}(2019)\citenamefont{Li, Sandhoefner, and
  Kovalev}}]{li2019intrinsic}
\bibinfo{author}{\bibfnamefont{B.}~\bibnamefont{Li}},
  \bibinfo{author}{\bibfnamefont{S.}~\bibnamefont{Sandhoefner}},
  \bibnamefont{and} \bibinfo{author}{\bibfnamefont{A.~A.}
  \bibnamefont{Kovalev}}, \bibinfo{journal}{arXiv preprint arXiv:1907.10567}
  (\bibinfo{year}{2019}).

\bibitem[{sup()}]{supplement}
\bibinfo{journal}{Supplementary Information}.

\bibitem[{\citenamefont{Shi et~al.}(2006)\citenamefont{Shi, Zhang, Xiao, and
  Niu}}]{shi2006proper}
\bibinfo{author}{\bibfnamefont{J.}~\bibnamefont{Shi}},
  \bibinfo{author}{\bibfnamefont{P.}~\bibnamefont{Zhang}},
  \bibinfo{author}{\bibfnamefont{D.}~\bibnamefont{Xiao}}, \bibnamefont{and}
  \bibinfo{author}{\bibfnamefont{Q.}~\bibnamefont{Niu}},
  \bibinfo{journal}{Phys. Rev. Lett.} \textbf{\bibinfo{volume}{96}},
  \bibinfo{pages}{076604} (\bibinfo{year}{2006}).

\bibitem[{\citenamefont{Ashcroft and Mermin}(1976)}]{ashcroft1976solid}
\bibinfo{author}{\bibfnamefont{N.~W.} \bibnamefont{Ashcroft}} \bibnamefont{and}
  \bibinfo{author}{\bibfnamefont{N.~D.} \bibnamefont{Mermin}},
  \emph{\bibinfo{title}{Solid State Physics}} (\bibinfo{publisher}{New York:
  Holt, Rinehart and Winston}, \bibinfo{year}{1976}).

\bibitem[{\citenamefont{Mook et~al.}(2019)\citenamefont{Mook, Neumann, Henk,
  and Mertig}}]{mook2019spin}
\bibinfo{author}{\bibfnamefont{A.}~\bibnamefont{Mook}},
  \bibinfo{author}{\bibfnamefont{R.~R.} \bibnamefont{Neumann}},
  \bibinfo{author}{\bibfnamefont{J.}~\bibnamefont{Henk}}, \bibnamefont{and}
  \bibinfo{author}{\bibfnamefont{I.}~\bibnamefont{Mertig}},
  \bibinfo{journal}{arXiv preprint arXiv:1903.11896}  (\bibinfo{year}{2019}).


\bibitem[{\citenamefont{Stamm et~al.}(2017)\citenamefont{Stamm, Murer,
  Berritta, Feng, Gabureac, Oppeneer, and Gambardella}}]{stamm2017magneto}
\bibinfo{author}{\bibfnamefont{C.}~\bibnamefont{Stamm}},
  \bibinfo{author}{\bibfnamefont{C.}~\bibnamefont{Murer}},
  \bibinfo{author}{\bibfnamefont{M.}~\bibnamefont{Berritta}},
  \bibinfo{author}{\bibfnamefont{J.}~\bibnamefont{Feng}},
  \bibinfo{author}{\bibfnamefont{M.}~\bibnamefont{Gabureac}},
  \bibinfo{author}{\bibfnamefont{P.~M.} \bibnamefont{Oppeneer}},
  \bibnamefont{and}
  \bibinfo{author}{\bibfnamefont{P.}~\bibnamefont{Gambardella}},
  \bibinfo{journal}{Phys. Rev. Lett.} \textbf{\bibinfo{volume}{119}},
  \bibinfo{pages}{087203} (\bibinfo{year}{2017}).
\bibitem[{\citenamefont{Togo et~al.}(2015)\citenamefont{Togo, Chaput, and
  Tanaka}}]{togo2015distributions}
\bibinfo{author}{\bibfnamefont{A.}~\bibnamefont{Togo}},
  \bibinfo{author}{\bibfnamefont{L.}~\bibnamefont{Chaput}}, \bibnamefont{and}
  \bibinfo{author}{\bibfnamefont{I.}~\bibnamefont{Tanaka}},
  \bibinfo{journal}{Phys. Rev. B} \textbf{\bibinfo{volume}{91}},
  \bibinfo{pages}{094306} (\bibinfo{year}{2015}).

\bibitem[{\citenamefont{Murakami et~al.}(2003)\citenamefont{Murakami, Nagaosa,
  and Zhang}}]{murakami2003spin}
\bibinfo{author}{\bibfnamefont{S.}~\bibnamefont{Murakami}},
  \bibinfo{author}{\bibfnamefont{N.}~\bibnamefont{Nagaosa}}, \bibnamefont{and}
  \bibinfo{author}{\bibfnamefont{S.-C.} \bibnamefont{Zhang}},
  \bibinfo{journal}{Phys. Rev. Lett.} \textbf{\bibinfo{volume}{93}},
  \bibinfo{pages}{156804} (\bibinfo{year}{2004}).

\bibitem[{\citenamefont{Canonico et~al.}(2019)\citenamefont{Canonico, Cysne, Rappoport, and
  Muniz}}]{canonico2019two}
\bibinfo{author}{\bibfnamefont{L.~M.}~\bibnamefont{Canonico}},
  \bibinfo{author}{\bibfnamefont{T.~P.}~\bibnamefont{Cysne}},
  \bibinfo{author}{\bibfnamefont{T.~G.}~\bibnamefont{Rappoport}}, \bibnamefont{and}
  \bibinfo{author}{\bibfnamefont{R.~B.} \bibnamefont{Muniz}},
  \bibinfo{journal}{arXiv preprint arXiv:1908.00927}  (\bibinfo{year}{2019}).
  
 \bibitem[{\citenamefont{Canonico et~al.}(2020)\citenamefont{Canonico, Cysne, Molina-Sanches, Muniz, Rappoport}}]{canonico2020orbital}
\bibinfo{author}{\bibfnamefont{L.~M.}~\bibnamefont{Canonico}},
  \bibinfo{author}{\bibfnamefont{T.~P.}~\bibnamefont{Cysne}},
  \bibinfo{author}{\bibfnamefont{A.}~\bibnamefont{Molina-Sanchez}},
  \bibinfo{author}{\bibfnamefont{R.~B.} \bibnamefont{Muniz}},
 \bibnamefont{and}
  \bibinfo{author}{\bibfnamefont{T.~G.}~\bibnamefont{Rappoport}}, 
  \bibinfo{journal}{arXiv preprint arXiv:2001.03592}  (\bibinfo{year}{2020}).
  
  \bibitem[{\citenamefont{Ziman}(1976)}]{ziman2001electrons}
\bibinfo{author}{\bibfnamefont{Z.~M.} \bibnamefont{Ziman}},
  \emph{\bibinfo{title}{Electrons and Phonons}} (\bibinfo{publisher}{Oxford University Press}, \bibinfo{year}2001).
\end{thebibliography}

\begin{thebibliography}{100}
\expandafter\ifx\csname natexlab\endcsname\relax\def\natexlab#1{#1}\fi
\expandafter\ifx\csname bibnamefont\endcsname\relax
  \def\bibnamefont#1{#1}\fi
\expandafter\ifx\csname bibfnamefont\endcsname\relax
  \def\bibfnamefont#1{#1}\fi
\expandafter\ifx\csname citenamefont\endcsname\relax
  \def\citenamefont#1{#1}\fi
\expandafter\ifx\csname url\endcsname\relax
  \def\url#1{\texttt{#1}}\fi
\expandafter\ifx\csname urlprefix\endcsname\relax\def\urlprefix{URL }\fi
\providecommand{\bibinfo}[2]{#2}
\providecommand{\eprint}[2][]{\url{#2}}

\bibitem[{\citenamefont{Luttinger}(1964)}]{luttinger1964theory}
\bibinfo{author}{\bibfnamefont{J.}~\bibnamefont{Luttinger}},
  \bibinfo{journal}{Phys. Rev.} \textbf{\bibinfo{volume}{135}},
  \bibinfo{pages}{A1505} (\bibinfo{year}{1964}).
  
 \bibitem[{\citenamefont{Shi et~al.}(2006)\citenamefont{Shi, Zhang, Xiao, and
  Niu}}]{shi2006proper}
\bibinfo{author}{\bibfnamefont{J.}~\bibnamefont{Shi}},
  \bibinfo{author}{\bibfnamefont{P.}~\bibnamefont{Zhang}},
  \bibinfo{author}{\bibfnamefont{D.}~\bibnamefont{Xiao}}, \bibnamefont{and}
  \bibinfo{author}{\bibfnamefont{Q.}~\bibnamefont{Niu}},
  \bibinfo{journal}{Phys. Rev. Lett.} \textbf{\bibinfo{volume}{96}},
  \bibinfo{pages}{076604} (\bibinfo{year}{2006}).
  
\bibitem[{\citenamefont{Matsumoto et~al.}(2014)\citenamefont{Matsumoto,
  Shindou, and Murakami}}]{matsumoto2014thermal}
\bibinfo{author}{\bibfnamefont{R.}~\bibnamefont{Matsumoto}},
  \bibinfo{author}{\bibfnamefont{R.}~\bibnamefont{Shindou}}, \bibnamefont{and}
  \bibinfo{author}{\bibfnamefont{S.}~\bibnamefont{Murakami}},
  \bibinfo{journal}{Phys. Rev. B} \textbf{\bibinfo{volume}{89}},
  \bibinfo{pages}{054420} (\bibinfo{year}{2014}).

\bibitem[{\citenamefont{Li et~al.}(2019)\citenamefont{Li, Sandhoefner, and
  Kovalev}}]{li2019intrinsic}
\bibinfo{author}{\bibfnamefont{B.}~\bibnamefont{Li}},
  \bibinfo{author}{\bibfnamefont{S.}~\bibnamefont{Sandhoefner}},
  \bibnamefont{and} \bibinfo{author}{\bibfnamefont{A.~A.}
  \bibnamefont{Kovalev}}, \bibinfo{journal}{arXiv preprint arXiv:1907.10567}
  (\bibinfo{year}{2019}).
  
  \bibitem[{\citenamefont{Hamada et~al.}(2018)\citenamefont{Hamada, Minamitani,
  Hirayama, and Murakami}}]{hamada2018phonon}
\bibinfo{author}{\bibfnamefont{M.}~\bibnamefont{Hamada}},
  \bibinfo{author}{\bibfnamefont{E.}~\bibnamefont{Minamitani}},
  \bibinfo{author}{\bibfnamefont{M.}~\bibnamefont{Hirayama}}, \bibnamefont{and}
  \bibinfo{author}{\bibfnamefont{S.}~\bibnamefont{Murakami}},
  \bibinfo{journal}{Phys. Rev. Lett.} \textbf{\bibinfo{volume}{121}},
  \bibinfo{pages}{175301} (\bibinfo{year}{2018}).
\end{thebibliography}
\end{document}